\documentclass[useAMS,usenatbib]{mn2e}
\usepackage{apjfonts}
\usepackage{graphicx}
\usepackage{epstopdf}
\usepackage{amssymb}
\newcommand{\apj} {ApJ}
\newcommand{\aj} {AJ}
\newcommand{\apjl} {ApJL}
\newcommand{\aap} {AAp}
\newcommand{\apjs} {ApJS}
\newcommand{\aaps} {AApS}
\newcommand{\mnras}{MNRAS}

\newcommand{\tab} {Table~}
\newcommand{\fig} {Figure~}
\newcommand{\sect} {Section~}

\title[The galaxy stellar  mass function at different local densities]{The importance of the local density in shaping the galaxy stellar mass functions\thanks{This paper includes data gathered with the 6.5 meter Magellan Telescopes located at Las Campanas Observatory, Chile}}

\author[Vulcani et al. ]{\parbox[t]{\textwidth}{Benedetta Vulcani$^{1,2}$\thanks{E-mail:
benedetta.vulcani@oapd.inaf.it }, Bianca M. Poggianti$^{2}$, Giovanni Fasano$^{2}$,  Vandana Desai$^3$, Alan Dressler$^{4}$, August Oemler Jr.$^{4}$,  Rosa Calvi$^1$, Mauro D'Onofrio$^1$ and Alessia Moretti$^{1,2}$    
}\\ 
\\
$^{1}$Astronomical Department, Padova University, Italy,\\ $^{2}$INAF-Astronomical Observatory of Padova, Italy,\\ 
$^3$Spitzer Science Center, California Institute of Technology, 
Pasadena, CA, USA\\
$^{4}$Observatories of the Carnegie Institution of Science,  Pasadena, CA, USA,\\
}

\begin{document}

\date{Accepted .... Received ..; in original form ...}

\pagerange{\pageref{firstpage}--\pageref{lastpage}} \pubyear{2011}

\maketitle

\label{firstpage}

\begin{abstract}
Exploiting the capabilities of four different surveys --- the
Padova-Millennium Galaxy and Group Catalogue (PM2GC), the WIde-field Nearby
Galaxy-cluster Survey (WINGS), the IMACS Cluster Building Survey (ICBS) and the ESO Distant Cluster Survey (EDisCS) --- we
analyze the galaxy stellar mass distribution as a function of local
density in mass-limited samples, in the field and in clusters from  low
($z\geq 0.04$) to high ($z\leq 0.8$) redshift.  We find that at all
redshifts and in all environments, local density plays a role in
shaping the mass distribution.  In the field, it regulates the
shape of the mass function at any mass above the mass limits. 
In clusters,
it seems to be important only at low masses ($\log M_{\ast}/M_{\odot} \leq
10.1$ in WINGS and $\log M_{\ast}/M_{\odot} \leq 10.4$ in EDisCS), otherwise 
it seems not to influence the mass distribution. 
Putting together our results with those of Calvi et al. and Vulcani et al. for the global environment, 
we argue that at least at $z\leq 0.8$ local density is more important than global 
environment in determining the galaxy stellar mass distribution, suggesting that
galaxy properties are not much dependent of halo
mass, but do depend on local scale processes.
\end{abstract}
\begin{keywords}
galaxies: clusters: general --- galaxies: evolution --- galaxies: formation --- galaxies:luminosity function, 
mass function --- galaxies: fundamental parameters
\end{keywords}

\section{Introduction}
It is well known that galaxies reside in environments that span a wide
range of galaxy densities (number of galaxies per Mpc$^3$).
Many authors have shown that galaxy density
plays an important role in determining many galaxy properties, such as
star formation rate, rest-frame colours, gas content and morphology
 (see, e.g., \citealt{dressler80, kauffmann04,baldry06,ellison09}).
Hence, if we wish to understand the physical processes that drive galaxy
evolution, we have to test for systematic differences between galaxies
in various environments.

In addition, it is equally well known that galaxies are characterized
by a wide range of total stellar masses.  Several works have shown that
mass is a crucial parameter in driving galaxy evolution
 and have claimed
that in some cases mass plays a more important role than the
environment in influencing galaxy properties (see
e.g. \citealt{peng10,ruth11a}).  
We note that to fully characterize the importance of the mass, 
it would be very interesting and important to have
the total galaxy mass (dark+luminous), but that is observationally 
challenging to achieve.
Hence, all of the cited
studies in this paper invetigate only the galaxy stellar mass, as tracer of the luminous galaxy matter.
Among others, \cite{kauffmann04} have shown
that at low-z, at fixed stellar mass, there is nearly no dependence of
structural properties like Sersic index or concentration parameter on
local galaxy density.
\cite{baldry06} have found that  the colour-mass and colour-concentration index
relations do not depend strongly on environment, 
while the fraction of galaxies on the red sequence depends
strongly on both stellar mass and environment.
\cite{mouhcine07} have found no dependence
of the relationship between galaxy stellar mass and
gas-phase oxygen abundance on local galaxy density. 
 At higher redshifts in zCOSMOS,
\cite{scodeggio09} observed a significant mass and 
optical colour segregation, in the sense that the median
value of the mass distribution is larger and the rest-frame 
optical colour is redder in regions of high galaxy density. However, 
considering only galaxies in a narrow range of stellar
mass, they no longer observed any significant colour segregation with density.

Trying to disentangle the contribution of the environment
and mass on the evolution, in order to quantify separately their importance, 
has been the aim of several works.
Studies by \cite{vanderwel08} showed that morphology and structure
are intrinsically different galaxy properties, and that they depend 
differently on galaxy mass and environment. Structure mainly depends on
galaxy mass whereas morphology mainly depends on environment.
\cite{ruth11a} found
that galaxy colour and the
fraction of blue galaxies depends very strongly on stellar mass at $0.4 < z < 1$,
while there is  only a weak
dependence on local density.  
This environmental influence is most visible in the colours of intermediate-mass galaxies
($10.5 <\log M_{\ast}/M_{\odot} < 11$), whereas colours of lower- and higher-mass galaxies remain largely
unchanged with redshift and environment. Fixing the stellar mass, the colour-density relation
almost disappears, while the colour-stellar mass relation is present at all local densities.
They also found a weak correlation between stellar mass 
and environment at intermediate redshifts. 
Restricting their analysis to a subsample of red galaxies,  \cite{moresco10} also
found that the colour distribution is not strongly dependent on
environment for all mass ranges, exhibiting only a weak trend 
such that galaxies in overdense regions 
are redder than galaxies in
underdense regions. On the other hand, the dependence on mass is far
more significant, with the average colours of massive galaxies being redder 
than low-mass galaxies.
\cite{ruth11b} found that galaxy colour strongly correlates with stellar mass, but it does not with local
density at fixed mass at all redshifts up to $z\sim 3$. 

Since mass and environment may be also strictly linked, it is important
to know how one depends on the
other and in particular to understand whether 
the stellar mass distribution, 
usually regarded as an intrisic property of a galaxy,
can be influenced by the environment, being
tightly coupled for example
to the depth of the halo potential and thus the halo mass.
Massive elliptical galaxies are often
found in the cores of galaxy clusters, or at high local
densities, while lower mass spirals are preferentially located in the outskirts
of large structures or in small groups. 
However, 
massive ellipticals are also found in the field
(e.g. \citealt{colbert01}), and low-mass galaxies with elliptical
morphology are found preferentially at high local densities
(e.g. \citealt{roberts07}).

Overall, it is still not fully clear 
how a galaxy's stellar mass depends on the environment
and how this dependence evolves with redshift.

The distribution of galaxy stellar masses is also of fundamental importance
for studying the assembly of galaxies over cosmic time. Estabilishing whether 
 the environment can regulate the mass
distribution could add an important piece in the puzzle of galaxy
evolution, claryfing the relation between these two
quantities.  Both estimating galaxy masses and defining and characterizing the environment
have their own uncertaintes and limitations. 
Furthermore, all the mass estimates
are strictly linked to the adopted Initial Mass Function (IMF). 
It is implicitly assumed that the IMF is universal, but it 
could be different for galaxies of all types (see e.g. \citealt{Guna11}). 
Moreover, the stellar mass can be model dependent (see \citealt{maraston05} vs. \citealt{bc03} models) and
again the choice of the model affects differently galaxies of different ages/metallicities. Results of different models can be controversial and lead to different findings. As a consequence, mass estimates are subject to systematic uncertainties (due also to
star formation history and metallicity variations) that are of the order of at least a factor two or more.

As far as the environment is concerned, definitions used to properly 
characterize it are mostly arbitrary. 
First of all, we have to distinguish between global and local environment: 
in the first case, according to the host halo mass, 
galaxies are commonly subdivided into e.g. superclusters, clusters, groups, 
field galaxies, voids, while in the second case
environment is described through the estimates of the local density, 
which can be 
parametrized in several ways, 
following different techniques.  For example, it is
possible to fix the metric aperture in which the number of neighbours
of a galaxy are counted or to measure the distance to the $n^{th}$ nearest
neighbor (with $n$ typically in the range 5-10).
Even if there is a sort of general correlation between global and local environments, 
as we will show also in this paper,
the two definitions of environments are not at all equivalent \citep{mulderw11}.

Focusing on local environment, 
galaxy densities also critically depend on how the sample is selected:
adopting a magnitude limited or a mass limited sample entails
a different selection of galaxies involved in the estimates of local density
and hence results can strongly change, according to the selection choices and the limits adopted
(see,  e.g., \citealt{wolf09}).

No matter how local density is parametrized,
the variation of the galaxy stellar mass distribution in regions of different density 
has been observed for mass limited samples
 both in the local Universe (see, e.g., \citealt{kauffmann04,baldry06})
and at higher redshifts  (see, e.g., \citealt{bundy06,scoville07,scodeggio09,bolzonella10}).
All the previous studies generally agree in finding that the mass distribution is regulated by local density.
Galaxies in lower and higher density regions show different mass distributions, in 
the sense that
lower density regions are proportionally more populated by lower mass galaxies. 
However, all of these studies considered a quite wide range of densities and moreover they mainly compared
the most extreme environments, to maximize the possible differences.
All of them considered general field data, without focusing especially in clusters, while
in this paper we make a first attempt to investigate the importance of the local density 
in regulating the mass distribution in different environments both at low and intermediate redshifts, also
considering separately the cluster environment.
To do this, we use the Padova-Millennium Galaxy and Group 
Catalogue (PM2GC) ($0.039<z<0.11$) \citep{rosa}, 
the Wide Nearby Galaxy Survey (WINGS) ($0.04<z<0.07$) \citep{fasano06}, the 
IMACS Cluster Building Survey (ICBS)  ($0.25<z<0.45$) (Oemler et al. 2012, in preparation)
 and the ESO Distant Cluster
Survey (EDisCS) ($0.4<z<0.8$) \citep{white05} datasets.

This paper is organized as follows: in \S 2, we present all the
datasets used, describing the galaxy samples. In \S 3.1, we
begin our analysis by showing how the mass distribution depends on local
density for $z\sim0$ field galaxies, and in \S 3.2 we focus
our attention only on galaxy clusters at similar redshifts. In \S 3.3 we move to higher redshift field galaxies, and, finally, 
in \S 3.4 show the results for clusters at $0.5<z<0.8$.
 We follow with a discussion in \S 4 and summarize our most important findings in \S 5.

Throughout this paper, we adopt ($H_0$, ${\Omega}_m$,
${\Omega}_{\lambda}$) = (70 $\rm km \, \, s^{-1} \, Mpc^{-1}$, 
0.3, 0.7) and a \cite{kr01} IMF, in the range
of mass 0.1-100 $M_{\odot}$.

\section{DATA AND GALAXY SAMPLES}
To characterize the mass function in different local density
conditions, we take advantage of four different data-sets that allow
us to analyze galaxies at different redshifts and in different global
environments.

In the following, we refer to ``general field'' (as in the case of the
PM2GC) when we consider a wide portion of sky, including all
environments.  In contrast, we refer to ``field'' (as in the case of
the ICBS) when we start from a cluster survey and we exclude cluster
members to consider a non-cluster sample.  We need to adopt these
definitions given the selection criteria of our surveys (see below).

\subsection{PM2GC}\label{pm}
To analyse galaxies in the general field in the local Universe we use
data from the PM2GC \citep{rosa}, a catalog of group, binary and single
galaxies at 0.03$\leq$z$\leq$0.11 
drawn from the Millennium Galaxy Catalog (MGC)
\citep{liske03}, a deep 
38 $deg^2$ B-band imaging and optical spectroscopic survey, 
which
provides a high quality, complete representation of the nearby galaxy
populations.

A detailed description of the MGC survey strategy, the photometric and
astrometric calibration, the object detection and classification
can be found in  \cite{liske03}, while
the selection and properties of the galaxy groups are described in \cite{rosa}.
For this paper, it is worth knowing that 
the PM2GC spectroscopic sample is essentially complete
to  $M_{B}<$-18.7,
 so there is no need to apply a statistical completeness
correction. 
Absolute B-band
magnitudes were obtained k-correcting the observed SE{\small XTRACTOR}
'BEST' magnitudes (MAGAUTO, except in crowded region where the ISOCOR
magnitude was used instead), corrected for Galactic extinction.

In the PM2GC sample there are 176 groups with at least three members\footnote{Within
these groups  
a few very massive groups  ($\sigma \geq 500 km \, s^{-1}$), 
comparable to clusters,  are included (see \citealt{rosa}).}
at $0.04\leq z\leq 0.1$, comprising in total 1057 galaxies,
representing 43\% of the general field population. The median redshift
and velocity dispersion of these groups are $z=0.0823$ and $\sigma =
192 km/s$, respectively. 88\% of the groups have less than ten
members, and 63\% have less than five members. Non-group galaxies have
been subdivided into ``binary'' systems of two bright close
companions, and ``single'' galaxies with no bright companion within 1500 $km/s$ and 0.5 $h^{-1} Mpc$.  The
binary and single catalogs contain 490 and 1141 galaxies,
respectively, at $0.03\leq z \leq 0.11$.  The general  field altogether
comprises 3210 galaxies at $0.03 \leq z \leq 0.11$ and includes all
group, binary and single galaxies as well as other galaxies that
belong to groups but are outside each group radial limits or
the redshift range for groups. 

Stellar masses are taken from \cite{rosa} and
were determined using the relation between
 $M/L_{B}$ and rest-frame $(B-V)$ colour, following \cite{bj01}
($ \log \frac{M}{L_B}= -0.51 +1.45(B-V)$), and then they were
 converted to a \cite{kr01} IMF (for details refer to \citealt{rosa}). 
The accuracy of the measured masses 
is $\sim$0.2-0.3 {\it dex}.
As discussed in \cite{rosa2}, the completeness
mass limit for the PM2GC sample is $\log M_{\ast}/M_{\odot}$=10.25.
Our choice to adopt a mass limit is dictated by the 
need 
to ensure completeness,
i.e. to include all galaxies more massive than the limit regardless
of their colour or type. To determine this limit, we
have computed the mass of an object whose observed magnitude is equal
to the faint magnitude limit of the survey, and whose colour
is the reddest colour of a galaxy at the highest redshift considered. 
With this selection, we are sure that our results will not be affected for
example by
the Malmquist Bias effect which leads to the 
preferential detection of intrinsically bright objects. This effect is instead very important
in magnitude limited samples, where galaxies below a certain brightness are neglected.

The projected local galaxy density is derived from the circular area
$A$ that, in projection on the sky, encloses the $N$ nearest galaxies
brighter than an absolute $V$ magnitude. The projected density is then
$\Sigma=N/A$ in number of galaxies per $Mpc^{2}$. %square megaparsec.  
For each
galaxy in the PM2GC survey, the local galaxy density has been computed
from the circular area (A$_{5}$) containing the 5 nearest projected
neighbours within $\pm 1000 \rm \, km \, s^{-1}$ from the galaxy and
with $M_V \leq -19.85$, which is the V absolute magnitude limit at
which the sample is spectroscopically complete.  

Due to the peculiar geometry of the area covered by the PM2GC survey
(a stripe 0.6deg$\times$73deg across the sky), when the local density
decreases, the circular area A$_{5}$ tends to overflow more and more
the survey coverage area, thus producing increasingly unreliable
estimates of the local density. To overcome this problem, in measuring
local densities we used the photometric and spectroscopic information
for all galaxies in the regions of sky around the MGC ($\pm
1.5^{\circ}$) from the Sloan Digital Sky Survey (SDSS,
\citealt{york00}) and the Two degree Field Galaxy Redshift Survey
(2dFGRS, \citealt{colless01}), that together yielded a highly complete
sample in the regions of interest.

Hereafter, we consider only PM2GC galaxies above the completeness
limit $\log M_{\ast}/M_{\odot}$=10.25.  In this way, our final PM2GC
sample consists of 1583 galaxies. 

\subsection{WINGS}\label{wi}
The WIde-field Nearby Galaxy-cluster Survey (WINGS) is
a multiwavelength photometric and spectroscopic survey of 77
galaxy clusters at 0.04 < z < 0.07 \citep{fasano06}. Clusters
were selected in the X-ray from the ROSAT Brightest Cluster
sample and its extension \citep{ebeling98, ebeling00} and the X-ray
Brightest Abell-type Cluster sample \citep{ebeling96}. 
WINGS has obtained wide-field optical photometry (BV)
for all 77 fields \citep{fasano06, varela09}, as
well as infrared (JK) photometry \citep{valentinuzzi09},
optical spectroscopy \citep{cava09}, and  U-band (Omizzolo et al. 2011, in prep.) 
for a subset of the WINGS clusters.

For WINGS we consider only spectroscopically confirmed members of 
21 clusters.
This is the subset of clusters that have a
spectroscopic completeness (the ratio of the number of spectra
yielding a redshift to the total number of galaxies in the photometric
catalog) higher than 50\%.  The clusters used in this analysis are
listed in \tab\ref{tab:wi.cl}.
We apply a
statistical correction to account for spectroscopic incompleteness. 
This is obtained by
weighting each galaxy by the inverse of the ratio of the number of
spectra yielding a redshift to the total number of galaxies in the
photometric catalog, in bins of 1 mag \citep{cava09}.

As for the PM2GC, galaxy 
stellar masses have been determined using the relation between
 $M/L_{B}$ and rest-frame $(B-V)$ colour proposed by \cite{bj01}.
The spectroscopic magnitude limit of the WINGS survey is
V=20, corresponding  to a mass limit $\log M_{\ast}/M_{\odot}$=9.8,
 above which the sample is unbiased. 
For a detailed description of the stellar
estimates see \cite{morph}.

For each spectroscopically confirmed cluster member in WINGS, the
local density has been computed from the circular area (A$_{10}$)
containing the 10 nearest projected neighbors in the photometric
catalog (with or without spectroscopic membership) whose V-band
absolute magnitude would be $M_{V}\leq -19.5$ if they were cluster
members. As we only want to count as neighbours the members of the
cluster, a statistical correction for field galaxy contamination has been applied
to the counts
using Table~5 in \cite{berta06}. In particular, since the field counts
in the area containing the 10 nearest neighbors are not integer
numbers, A$_{10}$ is obtained interpolating the two A$_n$ areas for
which the corrected counts (or the number of spectroscopic members, if
greater than them) are immediately lower and greater than 10.

A similar interpolation technique has been also used when the circular
area containing the 10 nearest neighbors is not fully covered by the
available data (galaxies at the edges of the WINGS field). In this
case, at increasing $n$ (and the corresponding area A$_n$), a coverage
factor has been evaluated as the ratio between the circular area and
the area actually covered by the observations. Then, the counts $n$
have been first multiplied for the corresponding coverage factors and
after corrected for the field counts (again always including
the spectroscopic members). Finally, as in the previous case, A$_{10}$
has been obtained interpolating the two A$_n$ areas for which the
corrected counts are immediately lower and greater than 10.

It is important to stress that the way local density
estimates are computed in WINGS and PM2GC are different and, as a consequence,
they are not directly comparable. 
WINGS is not spectroscopically complete, therefore it is not possible
to use a region within $\pm 1000$ km/s around each galaxy to count all
neighbours, as is done in PM2GC.  In WINGS, as usually done in
clusters, we try to use only cluster members as neighbours adopting a
statistical subtraction to remove the interlopers.  Moreover, since
clusters are highly populated, we can count 10 neighbours around each
galaxy.

In contrast, the PM2GC is spectroscopically
complete, but a ``membership'' to a structure is not meaningful, so we
have to define the region to count neighbours adopting a fixed
velocity distance. Finally, we use only the 5th
neighbour around each galaxy to avoid sampling too large a volume that
would include physically distant galaxies in other haloes.  

Given the necessarily different criteria for neighbours,
it is not possible to compare directly the local density estimates
in the two samples.

\begin{table}
\centering
\begin{tabular}{ccc}
\hline
cluster name & z & $\sigma$  		\\
	   &	& ($km \, s^{-1}$)		\\
\hline
A1069 & 0.0653 & 690$\pm$ 68\\
A119 & 0.0444 & 862$\pm$ 52 \\
A151 & 0.0532 & 760$\pm$55  \\
A1631a & 0.0461 & 640$\pm$33\\
A1644 &0.0467 & 1080$\pm$ 54\\
A2382 & 0.0641 & 888$\pm$ 54\\
A2399 & 0.0578 & 712$\pm$ 41\\
A2415& 0.0575 & 696$\pm$ 51 \\
A3128& 0.06 & 883$\pm$ 41   \\
A3158& 0.0593 & 1086$\pm$ 48\\
A3266& 0.0593 & 1368$\pm$ 60\\
A3376 &0.0461 & 779$\pm$ 49 \\
A3395 & 0.05 & 790$\pm$ 42  \\
A3490 & 0.0688 & 694$\pm$ 52\\
A3556 & 0.0479 & 558$\pm$ 37\\
A3560 & 0.0489  & 710$\pm$41\\
A3809 & 0.0627 & 563$\pm$ 40\\
A500 &0.0678  & 658$\pm$48   \\
A754 & 0.0547 & 1000$\pm$ 48\\
A957x & 0.0451 &710$\pm$ 53 \\
A970 & 0.0591 & 764$\pm$ 47 \\
\hline
\end{tabular}
\caption{List of WINGS clusters analysed in this paper and their redshift $z$, and velocity dispersion 
$\sigma$. \label{tab:wi.cl}  }
\end{table}

In WINGS, only galaxies with $\log M_{\ast}/M_{\odot} \geq 9.8$,
lying within 0.6$R_{200}$\footnote{$R_{200}$ 
 is defined as the radius delimiting a 
sphere with interior mean density 200 times the critical 
density of the Universe at that redshift,
and is commonly used as an approximation for the cluster 
virial radius. The $R_{200}$  values for our structures are computed 
from the velocity dispersions using the formula 
$$
R_{200}=1.73\frac{ \sigma}{1000 (km \, s^{-1})}\frac{1}{\sqrt{\Omega_{\Lambda}+\Omega_{0}(1+z)^{3}}}h^{-1}  \quad  (Mpc)
$$ 
} 
(the largest radius covered approximately in all clusters) are considered.
Moreover, Brightest Cluster Galaxies (BCGs) are excluded
from our analysis (see \citealt{morph} for details on the selection criteria).
The final WINGS sample consists of 1229 galaxies (1888 once weighted).

\subsection{ICBS} \label{ic}
The IMACS
Cluster Building Survey (ICBS) (Oemler et al. 2012, in preparation)
 is  a project focused on the study of
galaxy evolution and infall onto clusters from 
a clustercentric radius R $\sim$5 Mpc to the cluster regions.
Data have been acquired using the wide field of the 
Inamori-Magellan Areal Camera
and Spectrograph (IMACS) on Magellan-Baade.

The ICBS sought to define a homogeneous sample of clusters by  
selecting the most massive clusters per comoving volume at any  
redshift. Clusters were selected using the Red-Sequence Cluster Survey  
method \cite{gladders00}, either from the RCS  
itself, or from the Sloan Digital Sky Survey in regions of the sky not  
covered by the RCS. Within each field, galaxies were selected for  
observations from the RCS or SDSS catalogs, down to a limiting  
magnitude of $r \approx 22.5$. Spectroscopy of approximately 60\% of  
all objects brighter than this limit was obtained with the IMACS  
spectrograph on the 6.5m Baade Telescope at Las Campanas. Of those  
observed, only about 20\% failed to yield redshifts, or turned out to  
be stars. In addition, broad band photometry, in either the BVRI or  
griz systems, was obtained for each field, either with IMACS, or with  
the Wide Field CCD camera on the 2.5m duPont Telescope.

The data discussed in this paper come from four fields that contain
rich galaxy clusters at z =  0.33, 0.38, 0.42 and 0.43, as well as other 
structures at different redshifts. 
For this sample, we have decided to restrict our analysis to ICBS galaxies in the redshift range
$0.25<z<0.45$, in all the environments treated. This was done to
focus on a rather limited redshift range in order to use a common
mass limit set at $z=0.45$.
We treat separately cluster and field galaxies, hence we subdivide galaxies into two samples:
 ``clusters'' contain all galaxies within $\pm3 \sigma$ from the cluster redshift, the ``field'' 
include the others.

Since the projected density of cluster/supercluster members
is low at large clustercetric distances such as those probed by the ICBS, 
our sample  necessarily 
includes $\sim$1000 ``field'' galaxies at redshift $0.2 < z < 0.8$ per
survey field. 
The IMACS f/2 spectra have an observed-frame resolution of 10 \AA{}
full width at half-maximum with a typical $S/N\sim 20-30$ in the
continuum per resolution element. 

Details of the data and data analysis are presented in Oemler et al. 
(2012a, in preparation) and Oemler  et al. (2012b, in preparation).
Details on absolute magnitudes, mass estimates and completeness weights
can be found in \cite{global}. Briefly, 
absolute magnitudes have been determined using INTERREST
\citep{taylor09} from the observed photometry. 
When photometry is available, we determine the galaxy stellar
mass using the relation 
 between  $M/L_{B}$  and rest-frame $(B-V)$ colour
proposed by \cite{bj01}. The error of the measured masses
is $\sim$0.3 dex. As usual, all our masses are scaled to a \cite{kr01} IMF.
Our broadband photometry does not cover the entire field of our  
redshift survey. If photometry was not available for a galaxy,  
synthetic colours were calculated from the flux-calibrated IMACS spectra.

The magnitude completeness limit of the ICBS  is $r \sim
22.5$. At  our highest ICBS redshift, $z \sim 0.45$, 
%we determine the value of the
%mass of a galaxy with an absolute B magnitude corresponding to $r=22.5$,
%and a rest frame colour $(B-V) \sim 1$, which is the reddest colour 
%of galaxies in ICBS clusters.
%In this way, the ICBS 
the mass completeness limit %at the redshifts of interest
is $M_{\ast} = 10^{10.5} M_{\odot}$.

In this paper, galaxies are given weights proportional to the inverse of the
spectroscopic incompleteness.  Since the main galaxy property we wish to
analyse in this work is galaxy stellar mass, the incompleteness correction
has been computed
taking into account the number of
galaxies which have an estimate of the mass (for details, see \citealt{global}). 

Projected local densities  are derived from the rectangular area
$A$ that, in projection on the sky, encloses the $N$ nearest galaxies
brighter than $r=22.5$. The projected density is then
$\Sigma=N/A$ in number of galaxies per $Mpc^2$. 
Densities have been computed separately for cluster
and field galaxies and separately for each field. As a consequence,  local densities in the different global environments are 
not directly comparable. In both cases, 
local incompleteness has been taken into account. 
For clusters, local densities are derived taking into account all cluster
members and estimating the  area
 that encloses the 5  nearest galaxies.
For the field sample, densities
 are derived considering  5 nearest galaxies
 and within the rest frame velocity dispersion of $\pm 1000
km\,s^{-1}$. Due to the relatively small size of our fields, 
it has not been possible to find any companion within  $\pm 1000
km\,s^{-1}$ for some field galaxies. They are a ``very isolated sample'' and we gave to them a very low 
value of local density, so that they will be included in the lowest local density bin.

In the cluster sample, we exclude BCGs, whose properties could alter the general trends, 
 and consider all members regardless of their clustercentric distance.
The final mass-limited ICBS sample with 
$M_{\ast} \geq 10^{10.5} M_{\odot}$
consists of 371 galaxies.
Considering also the completeness weights, the number of galaxies is 754.
The field galaxy sample
consists of 275 galaxies, 658 once weighted.

\subsection{EDisCS}\label{ed}
For intermediate redshift clusters, we also use galaxies that belong to the 
ESO Distant Clusters Survey (EDisCS),
which is a multiwavelength photometric and spectroscopic survey 
of galaxies in 20 fields containing galaxy clusters at 
$0.4< z <1$ \citep{white05}. 
EDisCS clusters were drawn from the Las Campanas Distant Cluster Survey
(LCDCS) catalog \citep{gonzales01}. They were selected as surface
brightness peaks in smoothed images taken with a very wide optical
filter ($\sim 4500-7500$ \AA{}), 
and have high quality multiband optical and
near-IR photometry \citep{white05} and spectroscopy (\citealt{halliday04}, 
\citealt{milvang08}).

Photometric redshifts were computed for each object in the EDisCS
fields using two independent codes, a modified version of the publicly
available Hyperz code \citep{bolzonella00} and the code of
\cite{rudnick01} with the modifications presented in
\cite{rudnick03, rudnick09}. The accuracy of both methods is $\sigma (\delta z)
\sim 0.05-0.06$, where $\delta z = \frac{zspec-zphot} {1+zspec}.$
Photo-z membership (see also \citealt{delucia04} and \citealt{delucia07} for details)
 was established using a modified version of the technique
first developed in \cite{brunner00}, in which the probability of a
galaxy to be at redshift $z$ ($P(z)$) is integrated in a slice
$\Delta z = \pm 0.1$ around
the cluster redshift to give $P_{clust}$ for the two codes. 
A galaxy was rejected from the
membership list if $P_{clust}$ was smaller than a certain probability
$P_{thresh}$ for either code.  The $P_{thresh}$ value for each cluster
was calibrated from our spectroscopic redshifts and was chosen to
maximize the efficiency with which we can reject spectroscopic
non-members while retaining at least $\sim 90\%$ of the confirmed
cluster members, independent of their rest-frame (B-V) colour or
observed (V-I) colour. 

In \cite{vulcani10} we estimated galaxy stellar masses using photo-z
fitting total absolute magnitudes \citep{pello09} and, 
using, as for other surveys,  the relation between mass-to-light $M/L_B$ ratio and
rest-frame ($B-V$) colour for solar metallicity from \cite{bj01}.  The
photometric magnitude limit ($I=24$) corresponds to a mass limit $\log
M_{\ast}/M_{\odot}$= 10.2 \citep{vulcani10}.

The projected local galaxy density  is derived
from the circular area $A$ that in projection on the sky encloses
the $N$ closest galaxies brighter than an absolute $V$ magnitude
$M_V \leq =  -20$. 
We use $N= 10$, as in WINGS and in
most previous studies in clusters. Since for about only 7\% of the galaxies
in our sample the circular region containing the 10 nearest
neighbors extends off the chip and local densities only of these
sources suffer from edge effects, we do not use any interpolation technique, 
but simply excluded those galaxies from our
analysis.

As largely discussed in \cite{poggianti08},
we apply three different methods to identify the 10 cluster members
that are closest to each galaxy. These yield three different estimates
of the projected local density, which we compare in order to
assess the robustness of our results. Briefly, 
in the first method, the density is calculated using all galaxies
in our photometric catalogs and is then corrected using a statistical
background subtraction. In the other two methods we include only those galaxies that
are considered cluster members according to photometric redshift
estimates. 
We use two different criteria to identify photo-z neighbours. 
In the first case, a galaxy is accepted as neighbour
if it is a cluster member according to the photo-z membership criteria
described above.
In the other method
a galaxy is retained as neighbour if the best photometric
estimate of its redshift from the Hyperz code is within  0.1 in z
from the cluster redshift. 

Since all the methods give results that are in good agreement, in the following we show 
only the analysis of the second method, defining galaxy neighbours
according to photo-z membership using the integrated probability.

In this work, we use photo-z members of all the EDisCS clusters (see
Table \ref{tab:ed_cl} for the list of clusters) and we consider a mass
limited sample of galaxies with $\log M_{\ast}/M_{\odot}\geq  10.2$. We
take into account all cluster galaxies, regardless of their
clustercentric distance, but we exclude the BCGs, as we do in WINGS,
since their presence could alter the mass distributions. The final
EDisCS sample consists of 1560 galaxies.

\begin{table}
\centering
\begin{tabular}{|c|c|c|}
\hline
cluster name 	& z 	& $\sigma$  \\
	  	&	& ($km \, s^{-1}$)\\
\hline
cl 1018.8-1211 	&0.47	&486$^{+59}_{-63}$ \\
cl 1040.7-1155 	& 0.70 	&418$^{+55}_{-46}$\\
cl 1054.4-1146 	& 0.70 	&589$^{+78}_{-70}$\\ 
cl 1054.7-1245 	&0.75 	&504$^{+113}_{-65}$ \\
cl 1059.2-1253	&0.46	&510$^{+52}_{-56}$ \\
cl 1138.2-1133   	&0.48 	&732$^{+72}_{-76}$\\
cl 1202.7-1224	&0.42	&518$^{+92}_{-104}$\\
cl 1216.8-1201 	& 0.79 	&1018$^{+73}_{-77}$\\
cl 1227.9-1138	&0.64 	&574$^{+72}_{-75}$\\
cl 1232.5-1250 	& 0.54	&1080$^{+119}_{-89}$\\
cl 1301.7-1139	&0.48	&687$^{+81}_{-86}$\\
cl 1353.0-1137	&0.59	&666$^{+136}_{-139}$\\
cl 1354.2-1230    	& 0.76 	&648$^{+105}_{-110}$\\
cl 1411.1-1148	&0.52	&710$^{+125}_{-133}$\\
\hline
\end{tabular}
\caption{List of EDisCS clusters analysed in this paper, with cluster name, 
redshift $z$ and velocity dispersion $\sigma$ (from  \citealt{halliday04, milvang08}).
\label{tab:ed_cl}}
\end{table}

\section{RESULTS}
Above the mass completeness limit, we subdivide the galaxies of each
sample into four bins of local density,
so that in the two central bins galaxies are twice as numerous as
galaxies in the outer bins.
Since the choice of the number of
bins and of their limits is arbitrary, we tried also subdividing
galaxies into two, three, six, eight bins and checked that the final
conclusions are stable and independent from the choice made.

In each bin of local density, we build histograms to define the mass
distribution. In each mass bin, we sum all galaxies  to
obtain the total number of galaxies, then we divide this number 
by the width of the bin, to have the number of
galaxies per unit mass. The width of each mass bin is 0.2 dex.  
For building histograms of the WINGS and ICBS samples, each galaxy
 is weighted by its incompleteness
correction.
Errorbars on the $x-axis$ represent the width of the bin, errorbars on the
$y-axis$ are computed using poissonian errors \citep{gehrels86}.

In each of the following figures representing the mass functions, we have normalized the curves so that
the number of objects in the intermediate mass bins ($10.8\leq \log
M_{\ast}/M_{\odot} \leq 11.2$ ) is the same in all the mass functions
plotted.  In this way the differences at lower and higher masses are
easily visible at a glance.\footnote{We remind the reader that the normalization
adopted in displaying the mass functions doesn't influence the K-S
test and hence our results.}
We focus our attention mainly on 
the shape of the mass distribution.

With these aims, we use the  Kolmogorov-Smironov (K-S) test and also
visually analyse the plots.
The K-S test tells us whether we can disprove
the null hypothesis that two 
data sets are drawn from the same population distribution
function. The standard K-S, in building the
cumulative distribution, assigns to each object a weight equal to
1. Instead, our WINGS and ICBS data are characterized by spectroscopic
completeness weights.  So, when we study WINGS and ICBS galaxies, 
we modified the test, to make the relative
importance of each galaxy in the cumulative distribution depend on its
weight, and not being fixed to 1. 
A ``positive'' (statistically significant) K-S result robustly
highlights the differences between two distributions, but a negative
K-S result does not mean that the distributions are similar. In
particular, as we will see, when adopting low galaxy mass limits, the
K-S test is not sensitive to mass segregation at the high-mass end
simply because there are relatively few galaxies at high mass and they
are not able to sufficiently influence the cumulative distribution
upon which the K-S test is based.  Therefore it is necessary to
inspect the mass distributions, and their upper mass,
beyond the K-S test.

In the following, we present the results of our analysis for each
of our four galaxy samples. 
We start analysing the relation between
mass distribution and environment in the local Universe in the general
field, in order to consider a range of local density as large as
possible. Then, we 
focus our
attention only on clusters, to see if they behave as galaxies in the
general field. Subsequently, we move to higher redshift, where we 
again analyse both field and cluster galaxies.

We stress that we are not able to cross-compare our samples, 
at a given epoch or as a function of epoch. Since densities are defined 
in different ways for the each of the four samples 
(see \S\ref{pm}, \S\ref{wi}, \S\ref{ic}, \S\ref{ed}), the results of the  inter-sample 
analysis would be difficult to interpret.

\subsection{General field at low-z}
\begin{figure}
\centering
\includegraphics[scale=0.4]{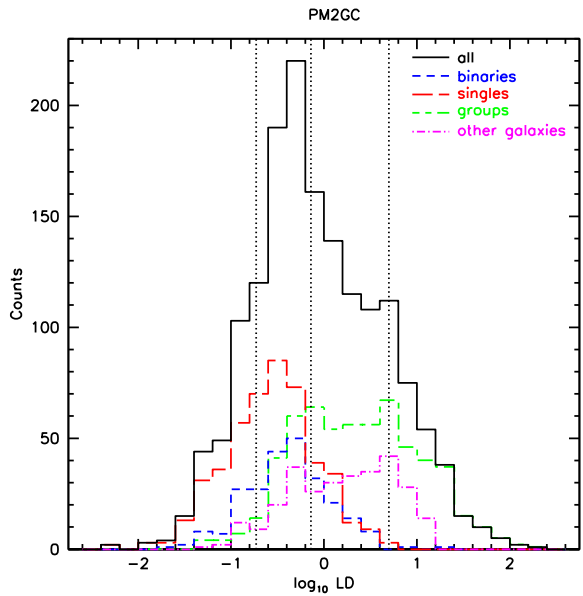}
\caption{Local density distribution of PM2GC galaxies at $z=0.3-0.11$
with $\log M_{\ast}/M_{\odot}\geq10.25$ for the whole field and
the different global environments. The vertical dotted lines represent 
the limits of our 4 density bins. 
\label{pm2gc_histo}}
\end{figure}

We use the PM2GC dataset to describe galaxies in the general 
field in the local Universe. 
Figure \ref{pm2gc_histo} shows the distribution of the local density
 in this sample and the limits adopted to 
subdivide galaxies into 4 bins. 
We can immediately see that the range of local densities spanned is
very wide, covering almost 4 dex.  
Even if  at first we will consider the general field altogether, 
it is useful to inspect the local density distributions of group, binary
and single galaxies separately, as shown
in \fig\ref{pm2gc_histo}.\footnote{For the sake of completeness, also ``other 
galaxies'' are plotted, they include all galaxies that belong to groups but are outside each
group radial limit or the redshift range for groups.}
Single galaxies are
preferentially located in the lowest density bins, groups in the highest
and binaries in the intermediate range. In particular, in
the lowest density bin the contribution of groups is almost
negligible, while in the highest bin single and binary galaxies are
almost absent.
Each environment, however, spans at least three of our density bins.

\begin{figure*}
\centering
\includegraphics[width=0.78\textwidth, angle=-90]{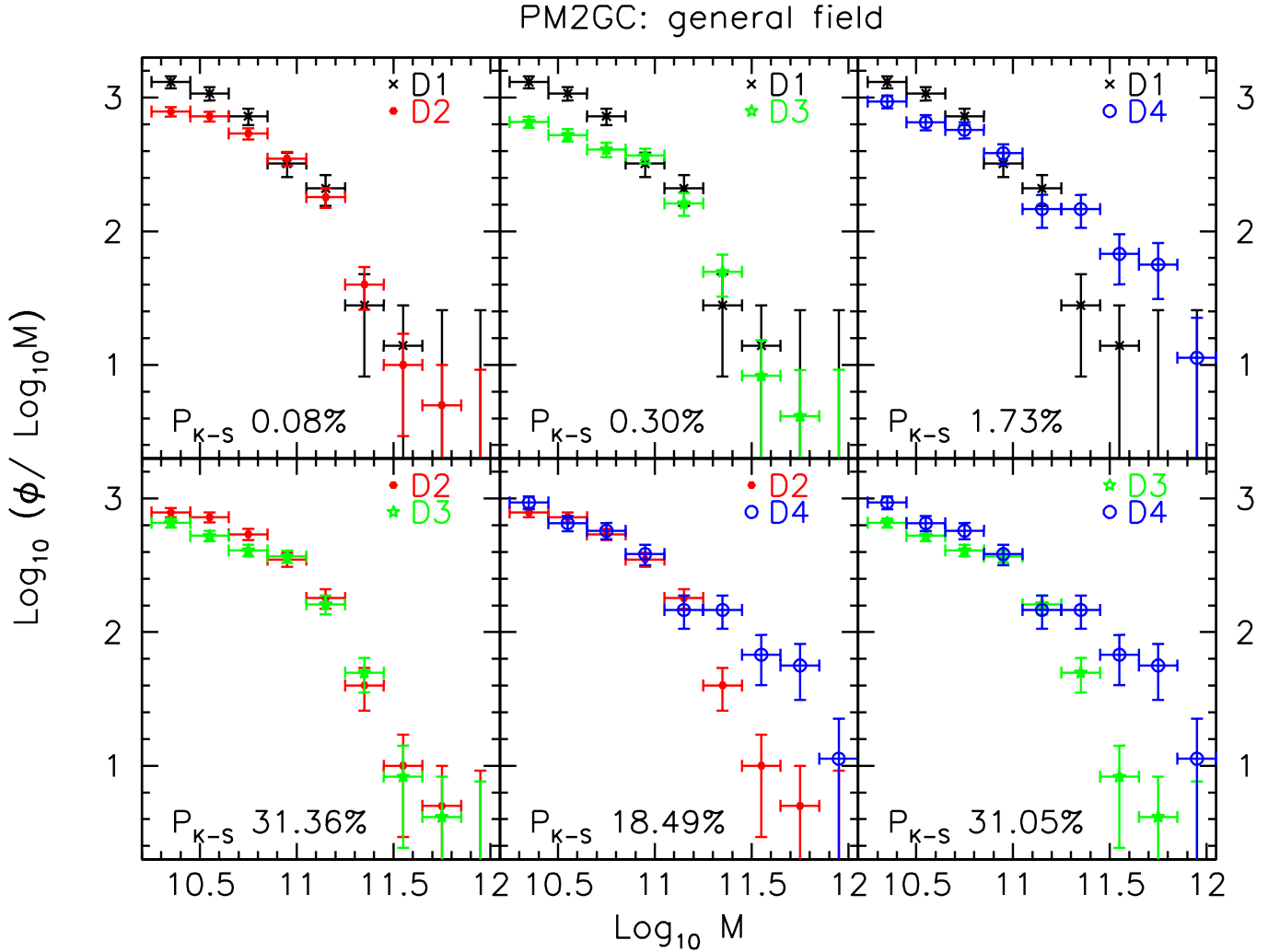}
\caption{Low-z  field (PM2GC): galaxy stellar mass functions in
four different bins of local density, compared two by two. The curves
are normalized so that the number of galaxies at intermediate masses
($10.8\leq \log M_{\ast}/M_{\odot} \leq 11.2$ ) is the same. Black
crosses: D1 (lowest density bin), red filled points: D2, green filled
stars: D3, blue empty points: D4 (highest density bin). Results of the
K-S test are also indicated. 
The mass function depends on local density: lower density
bins have proportionally a larger population of low mass galaxies than
higher density regions.
\label{pm2gc}}
\end{figure*}

In Figure \ref{pm2gc} we show the mass functions of galaxies in 
different density bins, compared two by two. 
We find that the mass function depends on local density: lower density
bins have proportionally a larger population of low mass galaxies than
higher density regions.

The K-S test can reject
the null hypothesis that the distributions are drawn from the same 
parent distribution
when we compare D1 with all the other bins, while it is inconclusive
in all other cases. 
However, looking at the figure, it clearly emerges that 
the slope of the D4 mass function at masses above $M_{\ast}/M_{\odot}\sim
10^{11}$
is much shallower than the slope in any other density bin, that
with the normalization adopted it is equivalent to say that in the highest 
density bin D4 there is an ``excess'' of high mass galaxies, compared to the 
other bins. 
To substantiate this on statistical grounds, 
since galaxies in the lowest mass bins are very numerous and they probably 
strongly influence the K-S test results, 
we try pushing up the mass
limit so to exclude those galaxies from the analysis. 
Redefining the mass limit entails a slightly change in the
limits of the local density bins, so we compute them
again.\footnote{From now on, when we change the mass limit, we always
compute again the limits of the density bins: each time, above the
adopted mass limit we subdivide galaxies so that in the two central
bins galaxies are twice as numerous as galaxies in the outer bins.}
For $ \log M_{\ast}/M_{\odot}\geq
10.5$, the differences in the mass function between D4 
and the other density bins become
statistically significant.

In general, even for a very high mass threshold ($ \log M_{\ast}/M_{\odot}\geq
10.8$), the differences in the mass functions of galaxies in different density bins remain
statistically significant, showing that local density matters
for any mass limit adopted.

As seen in \fig\ref{pm2gc_histo}, galaxies in groups, binary systems
and single galaxies cover different ranges of local densities,
therefore we now wish to test whether our local density results are
driven by galaxies in specific global environments (for example, only
in massive groups).  So we tried excluding single galaxies, or
galaxies located in massive groups ($\sigma_{group} >400 km/s$ and
$\sigma_{group} >500 km/s$) (plots not shown).  In all these cases we
always find a similar dependence of the mass functions on the local
density as we see in the general field.  Therefore, %we conclude that
the variations of the mass distributions with local density are not
driven by a different dependence in a specific  global environment.
 
\subsection{Clusters at low-z}

\begin{figure}
\centering
\includegraphics[scale=0.4]{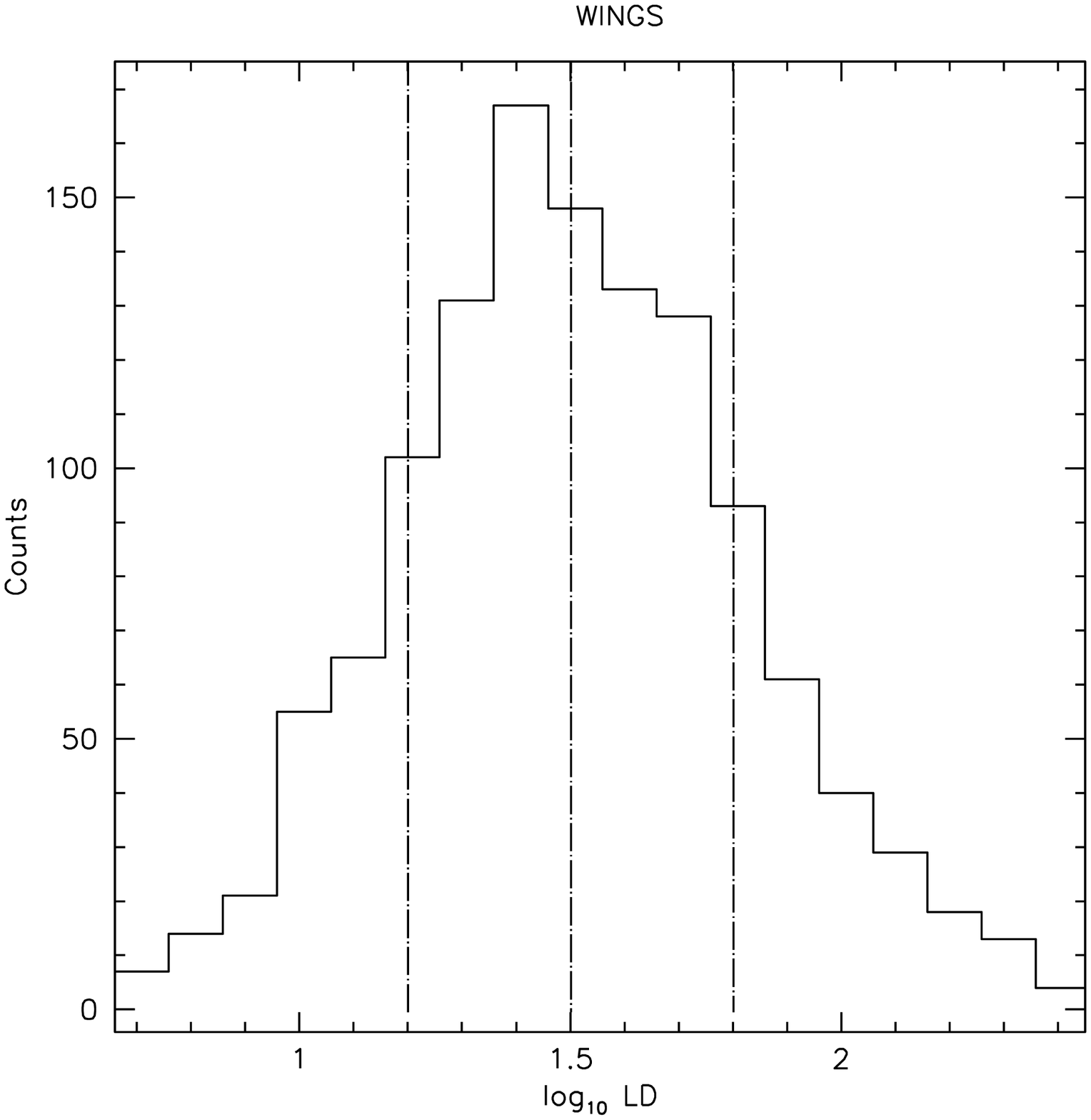}
\caption{Local density distribution of the WINGS sample for $\log M_{\ast}/M_{\odot}\geq9.8$. The vertical dotted lines represent 
the limits of our 4 density bins. \label{wings_histo}}
\end{figure}

We use the WINGS dataset to study galaxies in clusters in the local Universe.
Figure \ref{wings_histo} shows the distribution of local densities in
this sample and 
the 4 density 
bins. 
In clusters, galaxies
cover a range of local density of about 1.8 dex. 

\begin{figure*}
\centering
\includegraphics[width=0.78\textwidth, angle=-90]{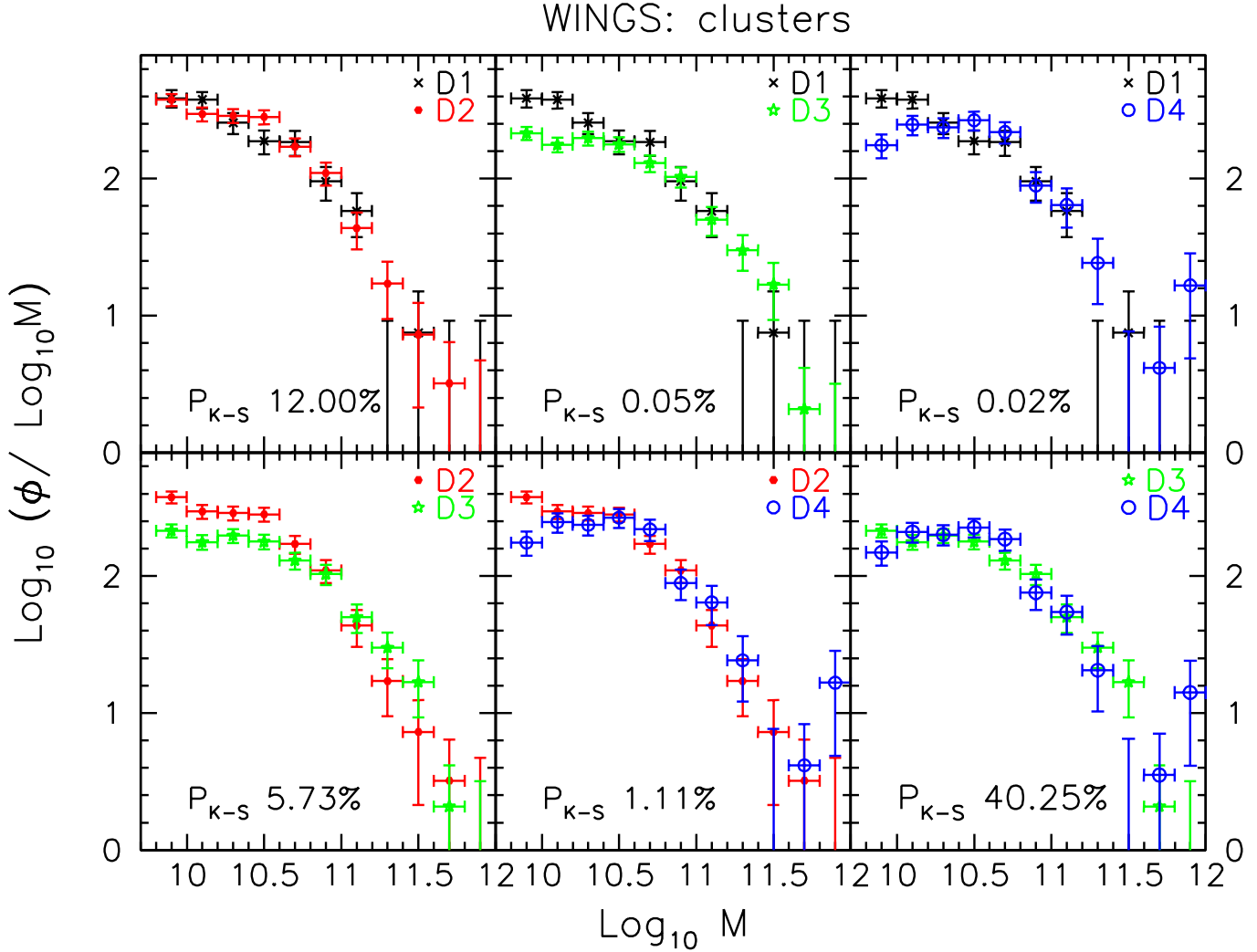}
\caption{WINGS: galaxy stellar mass functions in four different bins of local density, compared two by two. The curves are normalized so that the number of galaxies at intermediate masses ($10.8\leq \log M_{\ast}/M_{\odot} \leq 11.2$ ) is the same. Black crosses: D1, red filled points: D2, green filled stars: D3, blue empty points: D4.  Results of the K-S test are also indicated. As for the general field, the mass function depends on local density: in general, lower density
bins have proportionally a larger population of low mass galaxies than
higher density regions.
\label{wings}}
\end{figure*}

In Figure \ref{wings} we show the mass function of galaxies in the different bins of local density, compared two by two.\footnote{We remind the reader that the WINGS sample is not spectroscopically complete, so in all the following analysis, we always take into account WINGS' weights.} 
Also in the case of clusters,
there is a
dependence of the mass function on the local density. In general (except when
we compare D3 and D4 whose shapes are very similar),
lower density bins have proportionally a greater number of lower mass
galaxies. 
The
K-S test can reject the hypothesis of a common parent distribution
with a high level of significance in most of the cases.  

We note that, unlike the mass function in the general field at the same redshift, 
in clusters the mass functions in the highest density bin 
(D4, and perhaps D3) flattens out at low galaxy masses, below $\log M_{\ast}/M_{\odot}\sim 10.5$. This is
suggestive of a sort of ``deficit'' of low-mass galaxies with respect to intermediate
masses compared to lower density regions.

As before, the K-S test is particularly sensitive to the large number
of low mass galaxies, so we push up the mass limit in order to
detect possible differences in the slope of the mass functions at high
mass.
After redefining the density bins, we find that  local
density effects in low-z clusters are not visible at intermediate-high
galaxy masses, as the K-S finds differences between the mass function
of galaxies in different density regions only for a mass limit $\log
M_{\ast}/M_{\odot}\leq 10.1$.  This limit is even lower than the mass
completeness limit of the PM2GC survey.  In contrast, as we have seen
in \S3.1 for the PM2GC, the local density effects in the general field
on the shape of the mass function do not disappear at any mass.

\subsection{Field at intermediate-z}
\begin{figure}
\centering
\includegraphics[scale=0.4]{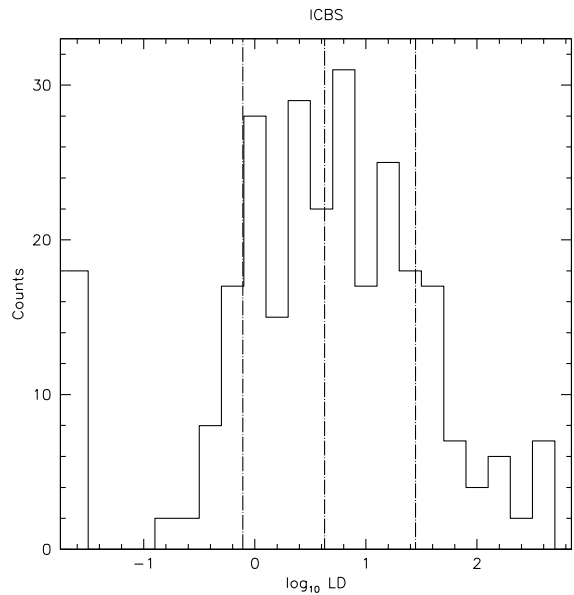}
\caption{Local density distribution of ICBSgalaxies at $z=0.25-0.45$
with $\log M_{\ast}/M_{\odot}\geq10.5$ for the field. Very isolated galaxies without an 
estimate of local density
(see \ref{ic}) are assigned $\log(LD)= -1.5$.
The vertical dotted lines represent 
the limits of our 4 density bins. 
\label{icbs_histo}}
\end{figure}

We use the field sample of the ICBS dataset to characterize galaxies in the 
field at intermediate redshifts. 
Figure \ref{icbs_histo} shows the distribution of the local density
 in this sample %and the limits adopted to 
%subdivide galaxies into 
and the 4 density bins. In the histogram, very isolated galaxies without an estimates of local density
(see \sect\ref{ic}) are assigned $\log(LD)= -1.5$.
We can immediately see that, excluding very isolated galaxies, for which we do not have a real
estimates of local density,  the range of local densities spanned is
very wide, covering almost 4 dex. This range is also very similar to that we found for the PM2GC, indicating that actually
the (general) field is a very  heterogeneous environment, with very sparse regions but also with highly populated ones.

\begin{figure*}
\centering
\includegraphics[width=0.78\textwidth, angle=-90]{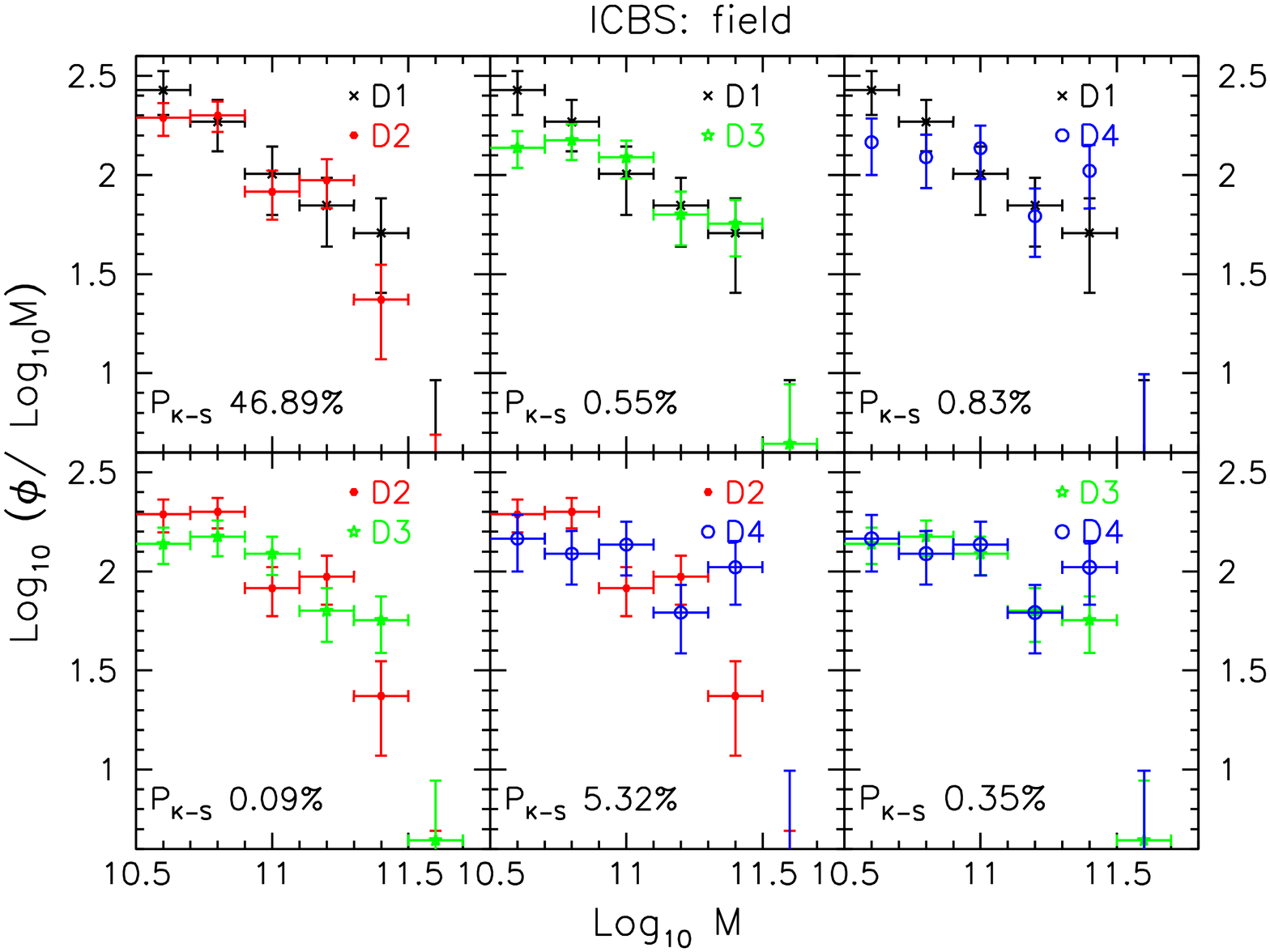}
\caption{Intermediate-z  field (ICBS): galaxy stellar mass functions in
four different bins of local density, compared two by two. The curves
are normalized so that the number of galaxies at intermediate masses
($10.8\leq \log M_{\ast}/M_{\odot} \leq 11.2$ ) is the same. Black
crosses: D1 (lowest density bin), red filled points: D2, green filled
stars: D3, blue empty points: D4 (highest density bin). Results of the
K-S test are also indicated. The mass function depends on local density: 
lower density
regions have proportionally a larger population of low mass galaxies than
higher density regions.
\label{icbs}}
\end{figure*}

In Figure \ref{icbs} we present the mass functions of galaxies in 
different density bins, compared two by two. 
Again, the mass function depends on local density in the sense that 
lower density
regions have proportionally a larger population of low mass galaxies than
higher density regions.

Despite the quite small number statistic, the K-S test can always reject
the null hypothesis that the distributions are drawn from the same 
parent distribution
except when we compare D1 and D2. 
Moreover, looking at the figure, as in the PM2GC, we find that 
the slope of the D4 (and maybe D3)  mass function at masses above $M_{\ast}/M_{\odot}\sim
10^{11.2}$
is shallower than the slope in the other density bins, indicating a possible 
``excess'' of high mass galaxies in that bin, compared to the 
other bins. 

In this case we decide not to further push up the mass limit, both because it is already fairly high 
and because the statistical uncertainty would be too large.

\subsection{Clusters at intemediate-z}
We use the  EDisCS dataset to describe galaxies in distant 
clusters. 
\begin{figure}
\centering
\includegraphics[scale=0.4]
{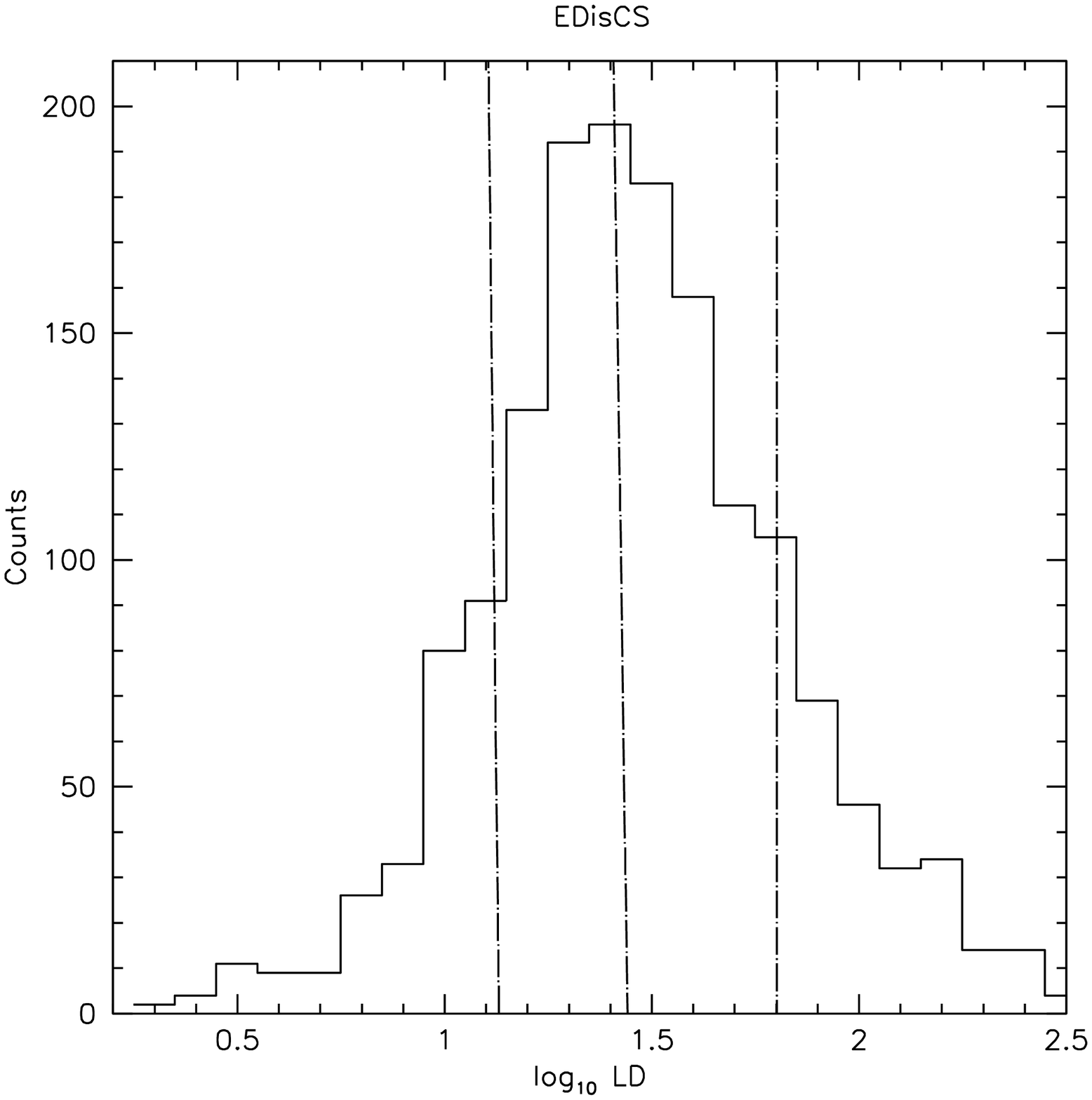}
\caption{Local density distribution of EDisCS for the 
sample with $\log M_{\ast}/M_{\odot}\geq10.2$.  
The vertical dotted lines represent 
the limits of our 4 density bins.
\label{ediscs_histo} }
\end{figure}

Figure \ref{ediscs_histo} shows the distribution of the local density 
in this sample and the limits % adopted to 
%subdivide galaxies into 
of the 4 density bins. 
In clusters at high-z, galaxies can assume local density values in a range of almost
2.3 dex. 

\begin{figure*}
\centering
\includegraphics[width=0.78\textwidth, angle=-90]{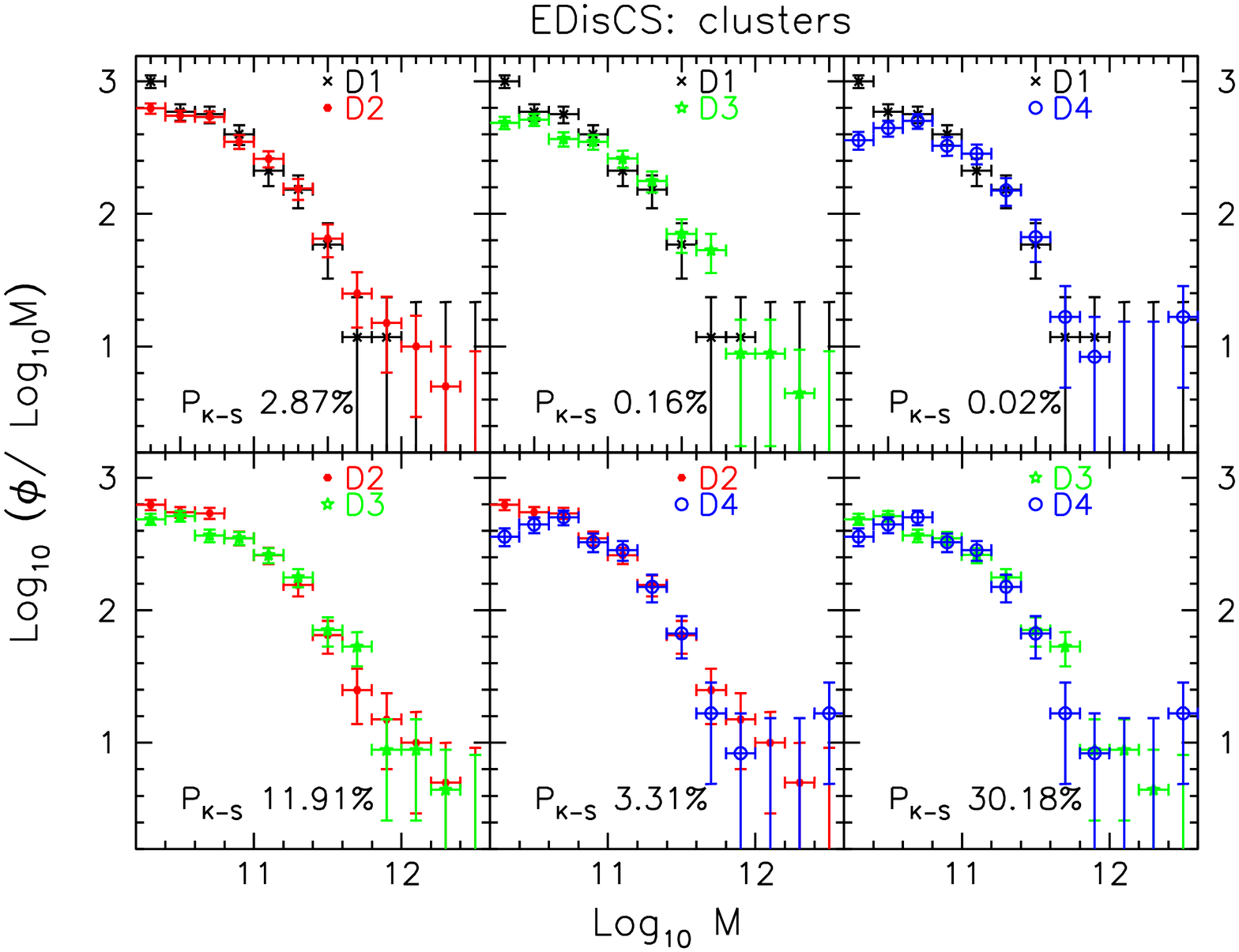}
\caption{EDisCS: galaxy stellar mass functions in four different bins of local density, compared two by two. The curves are normalized so that the number of galaxies at intermediate masses ($10.8\leq \log M_{\ast}/M_{\odot} \leq 11.2$ ) is the same. Black crosses: D1, red filled points: D2, green filled stars: D3, blue empty points: D4.  Results of the K-S test are also indicated. As for clusters in the local Universe, there is a clear dependence of the
 mass function on the local density: low density
 regions have proportionally more low mass galaxies. \label{ediscs}}
\end{figure*}

In Figure \ref{ediscs} we show the mass functions of galaxies in the different bins, compared two by two. 
As for WINGS, we find that there is a clear dependence of the
 mass function on the local density, in the sense that low density
 regions have proportionally more low mass galaxies. 
The K-S test is 
able to refuse the null hypothesis of similarity of the populations 
in the majority of cases.

Again, as for WINGS, also in distant clusters the mass functions in the highest density bin 
(D4, and perhaps D3) flattens out at low galaxy masses, 

Adopting a higher mass limit, we find that for $\log
M_{\ast}/M_{\odot}\geq 10.4$ the K-S results are inconclusive,
suggesting again that in clusters the effects of the local density
on the high mass end shape of the mass function are not visible.

Using the ICBS cluster sample (plots not shown), above the mass limit $ \log M_{\ast}/M_{\odot}>10.5$
we find that the local density range spanned is $\sim$ 2.5 dex, very similar to
that of EDisCS. Moreover, 
in agreement with the EDisCS
findings, we find no dependence at such high masses.

\section{General trends}
Our results show that at both redshifts and in all environments
there is a dependence of the mass function on the local density. 
Even if we can not perform any inter-sample comparison, since densities have been
computed using different criteria, in the following we can qualitatively
compare our results coming from the different surveys, to detect if a common trend
does exist.

In general, the lower the density, the higher is (proportionally) the
number of low mass galaxies, indicating that low-mass galaxies are
more common in the ``sparsest'' regions.  

\begin{figure*}
\centering
\includegraphics[width=0.75\textwidth]{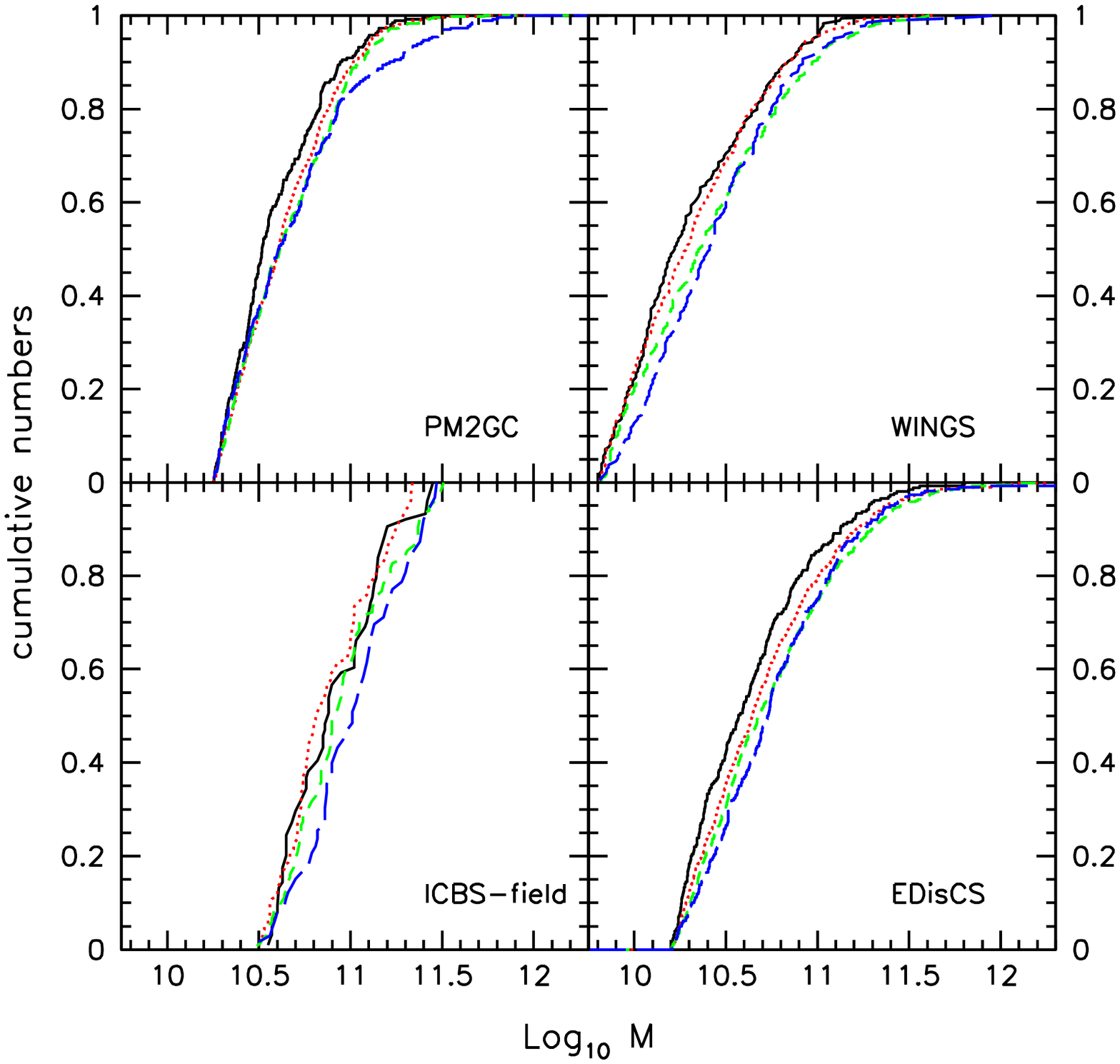}
\caption{Cumulative distributions of PM2GC (top left panel), 
WINGS (top right panel) and EDisCS (bottom right panel). In all panels, 
black solid lines represent the lowest density bin D1, red dotted lines D2, 
green dashed lines D3 and blue long dashed lines D4. In all sample, the lower
the density, the higher is proportionally the number of low mass galaxies. In all samples, also the highest masses reached at
different densities are different. 
 \label{cumulate}}
\end{figure*}

\fig\ref{cumulate} shows the cumulative distributions of PM2GC, WINGS, ICBS-field and EDisCS
and summarizes  the main result: 
in higher density regions, galaxies are proportionately more massive, indicating
that the mass function shape changes with local density (the colour progression black (D1), red (D2), green (D3), blue (D4) is always the same). 

Galaxies in D1, D2, D3 and D4 reach a different upper mass.
In particular, very massive
galaxies (having excluded the cluster BCGs) seem to be located only in the highest density bin, while they are
absent at lower densities. 
For example, in the PM2GC, 
the most massive galaxy in D1 has $\log M_{\ast}/M_{\odot}=11.5$, 
the most massive galaxy in D2 and D3 has $\log M_{\ast}/M_{\odot}=11.7$, while the 
 most massive galaxy in D4 has $\log M_{\ast}/M_{\odot}=11.9$.
In WINGS  neither D1, D2 nor D3 host galaxies more massive than 
$\log M_{\ast}/M_{\odot}=11.6$, while D4 is
populated also by galaxies with $11.6\leq \log
M_{\ast}/M_{\odot}\leq 12$. 
 This supports 
supporting the mass segregation scenario
for the very most massive galaxies.

\begin{figure*}
\centering
\includegraphics[scale=0.2]{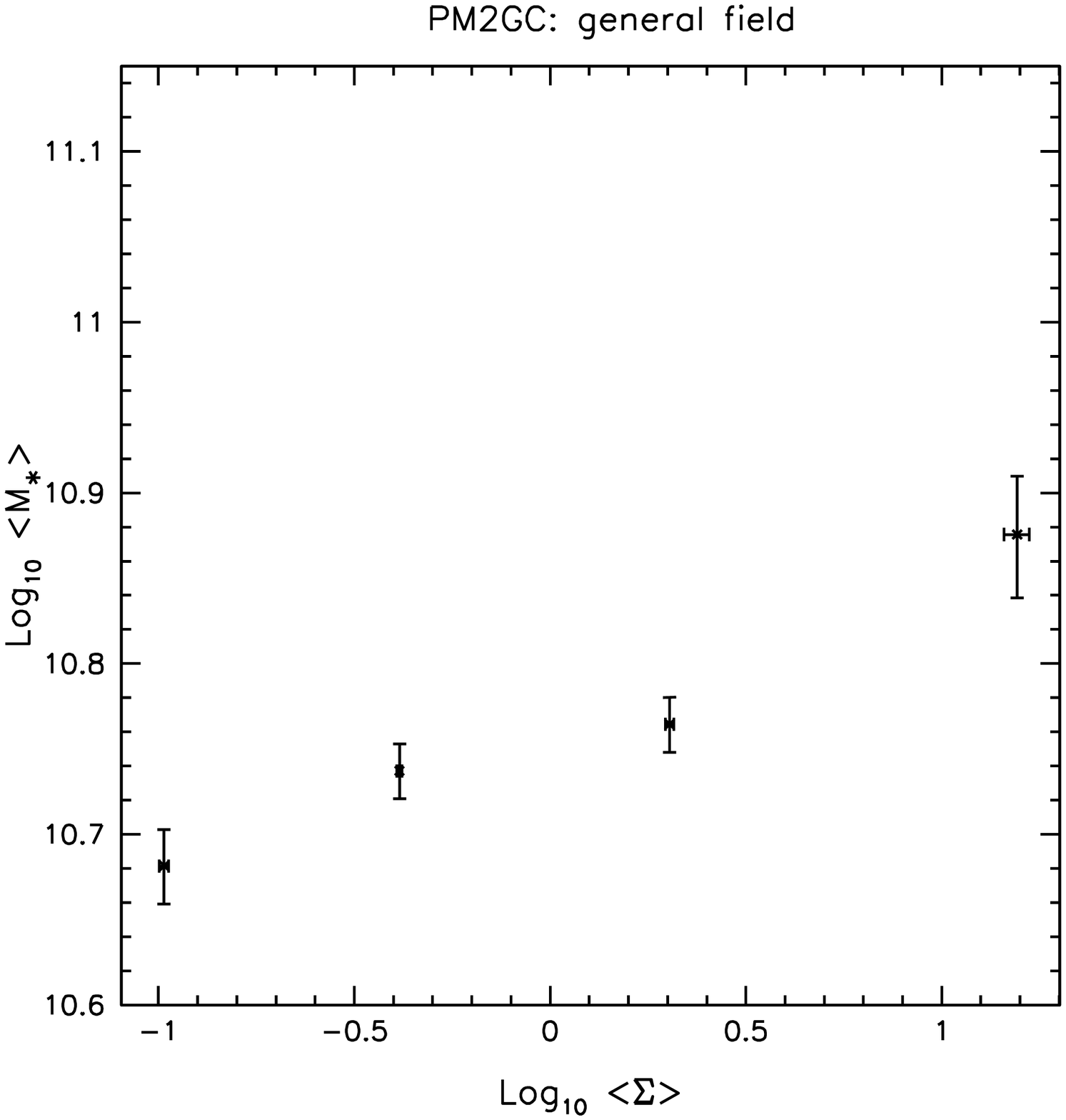}
\includegraphics[scale=0.2]{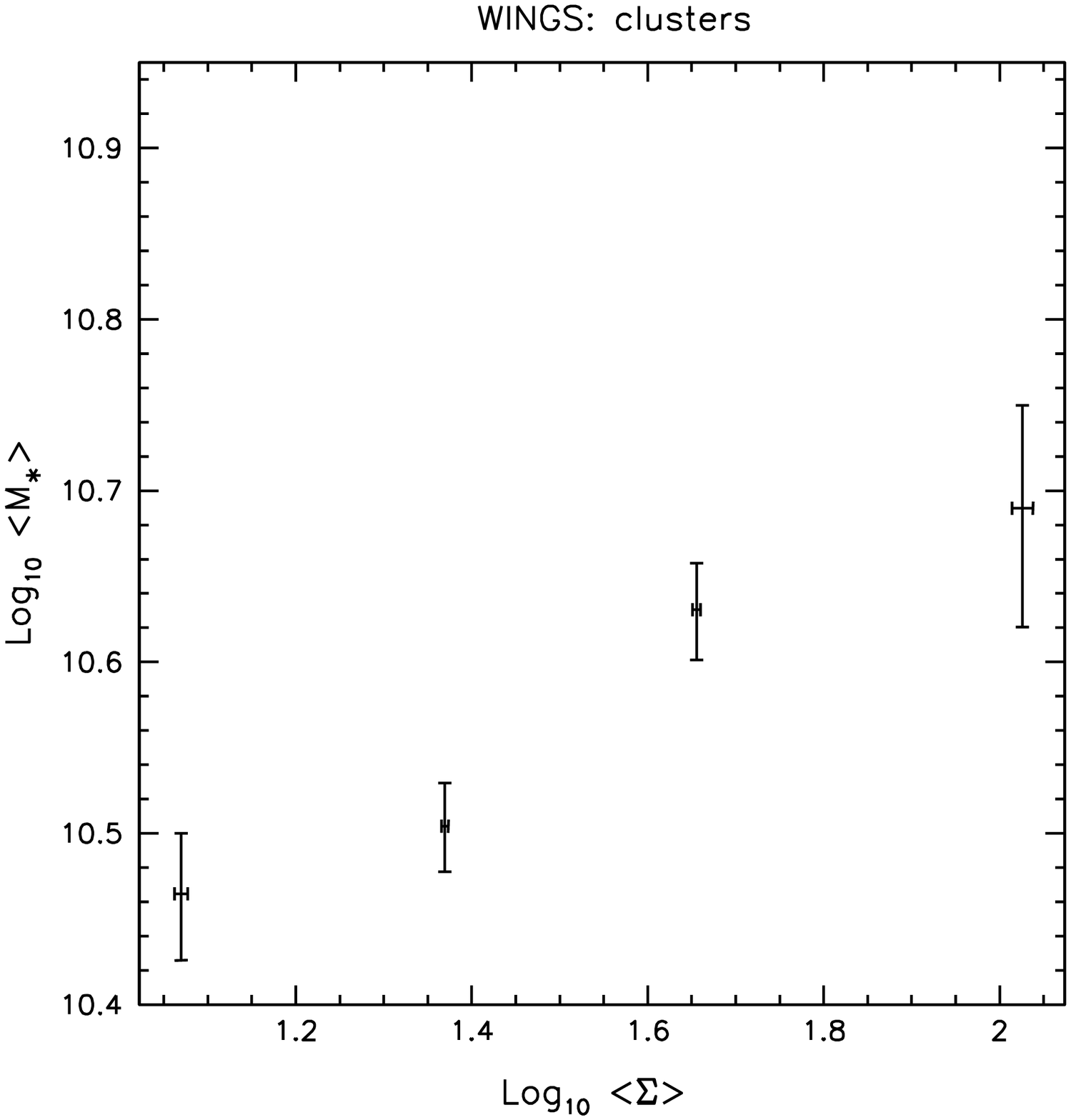}
\includegraphics[scale=0.2]{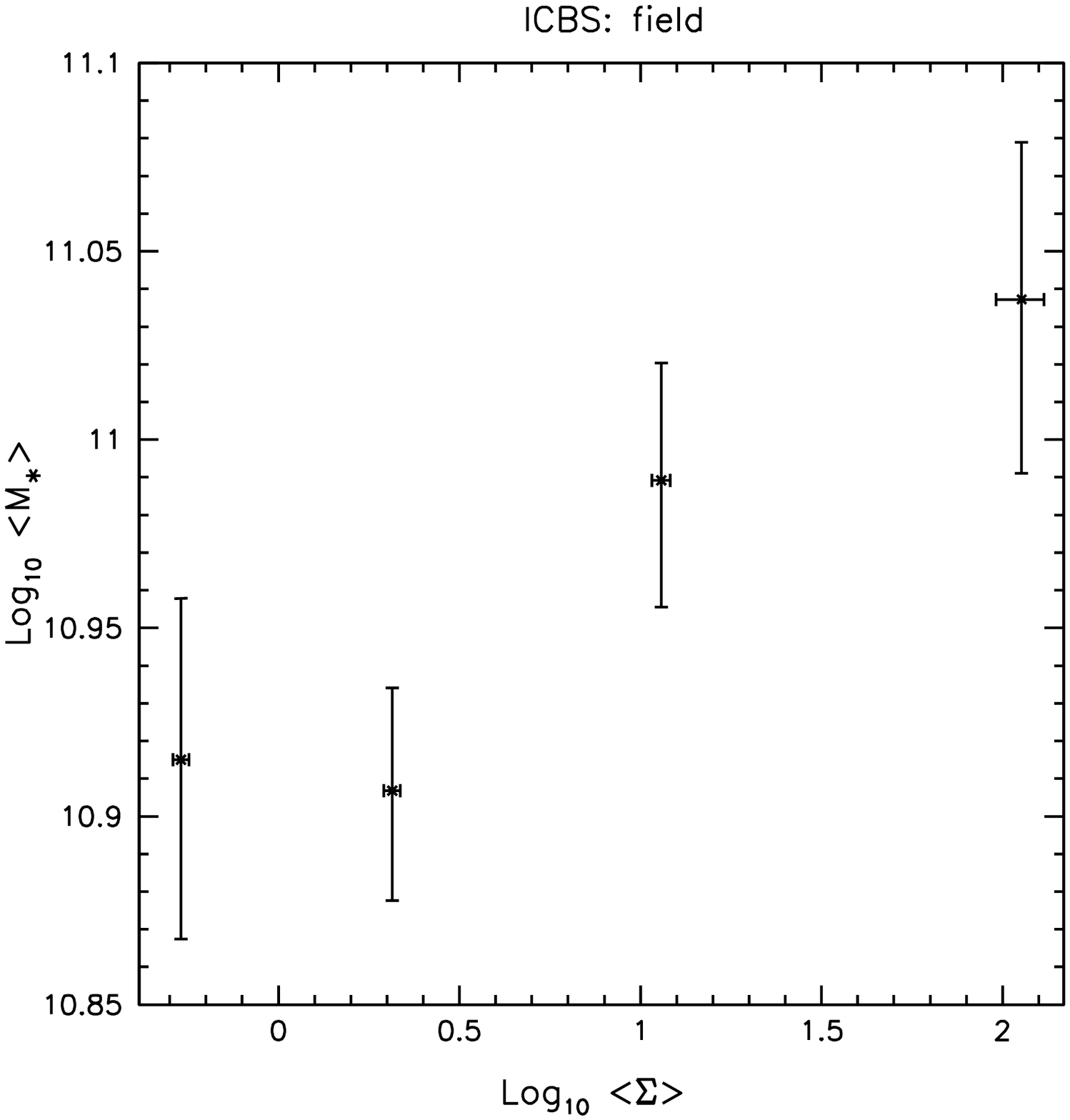}
\includegraphics[scale=0.2]{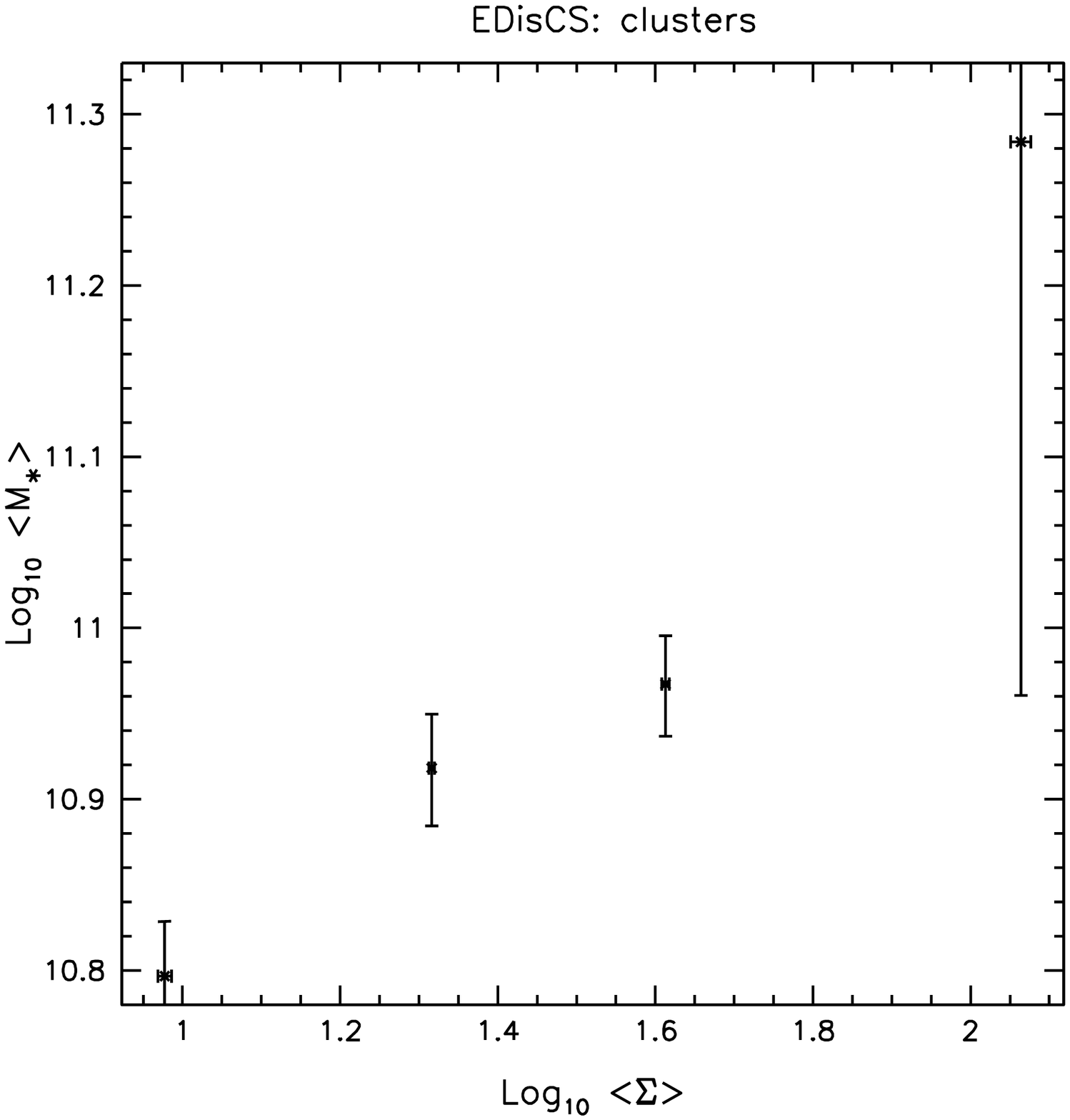}
\caption{Mass - local density relation for PM2GC (left panel), WINGS (central left panel), ICBS (central right panel) and EDisCS (right panel)
 surveys. For each sample, above 
its proper mass completeness limt, the mean mass has been computed separately
in the four density bins. Errors are defined as $rms / \sqrt{N}$ where $N$ is the number of galaxies in each density bin.
For WINGS and ICBS, weighted means are computed. In all samples, the average mass depends on local density: the average mass is higher
in higher density bins. 
\label{Md}}
\end{figure*}

  Not only the maximum mass depends on density, 
but also the average mass does, as shown in \fig\ref{Md} separately for each sample. In the figure, the
logarithmic mean mass computed in each density bin is plotted vs. the logarithmic mean density. 
The average mass allows us to have an immediate
comparison among the different characteristic masses at the different
local densities, and  to see how the
mean mass changes as a function of the LD in each sample.
 In all samples, only
galaxies above the mass limit of the sample  are considered, hence mean masses
can not be directly compared in the different samples.
We find a common trend in all samples: as it might be expected based on the results shown before, the average mass is higher
in higher density bins. In the local Universe, both in the  field and in clusters, 
the $\Delta (\log <M>)$ is about 0.2 dex, at intermediate redshift in the  field it is slightly higher 
($\Delta (\log <M>)\sim 0.25$ dex) while in distant clusters the difference between the mean mass in the lowest 
and the highest denisty bin is greater ($\Delta (\log <M>) \sim 0.5$ dex): on average, 
galaxies in D4 are much more massive than in other bins. 

Moreover, \fig\ref{cumulate} shows that galaxies more massive than
$\log M_{\ast}/M_{\odot} \sim 11$, (that represent 22\%, 22\% 37\% and 36\% of all galaxies more
massive than $\log M_{\ast}/M_{\odot}=10.5$ in PM2GC, WINGS ICBS and EDisCS, respectively\footnote{The fact that at low and intermediate redshift we find the same fraction of massive galaxies indicates that the evolution of the fraction is independent of environment.}) 
are not
confined to the highest density regions: about 20\% of them are in D4,
and about 70\% in D2+D3, in all the samples. 

So far, in the literature, several works analysed galaxies located in
regions characterized by different densities (all of them
for galaxies in the general field). For the local Universe, \cite{kauffmann04} and
\cite{baldry06} focused their attention on a wide range of local densities, while 
at higher redshift e.g. \cite{bundy06} and \cite{bolzonella10} focused
mainly on the extreme environments, usually comparing D1 and D4,
neglecting intermediate regions.  
Our results are in agreement with them. 
In addition, in most cases
we find differences also in the mass function of galaxies in contiguous density
bins.

\section{Global and local environments}
We have seen that (general) field and clusters seem to qualitatively 
behave in a quite
different way.  In the local general field, 
the local density can influence the stellar mass function at any mass:
comparing the mass function in different density bins, 
differences in mass function slope
are visible both in the low mass regime and at the high mass end.
In the higher redshift field, local density influences the
mass distribution  at high masses, but the rather high mass limit  
does not allow us to inspect lower masses. 
On the other hand, in clusters, the biggest differences are always confined
at low masses, while the shape of the 
mass function of intermediate-massive galaxies seems not
to be strongly affected by the local density. 

In principle, this different behaviour observed in the field and
clusters could be due to two different reasons: the smallest local
density range sampled in clusters, or a residual dependence on the
global environment.

In the local Universe, 
since the density range investigated with WINGS is
relatively small ($\Delta (\log <\Sigma>) =1.8dex$), it is possible that this range
of densities corresponds to only the highest density regions in the
PM2GC, that spans 
$\Delta (\log <\Sigma>) =4dex$. 
Unfortunately, there is no way to directly compare the local densities
in the two samples to surely assess how the density ranges overlap.
In any case, since the PM2GC contains also high velocity dispersion
structures, we have
analysed their density distribution separately. First of all,  
we have checked that the PM2GC "clusters"
(groups with $\sigma_{group} >500 km/s$) cover a range of local density
very similar to that spanned by {\it all} groups, 
hence their galaxies are located also in low
density regions (D1 and D2). Second, we have checked
the host structure of galaxies in the highest density bin D4 and we
have found that actually only 43\% of galaxies in D4 belong to a
structure with a velocity dispersion $\sigma>400 km/s$ and 30\% belong
to a structure with $\sigma >500 km/s$. Therefore, D4 is also
populated by galaxies in smaller systems not comparable to the cluster
environment.  On the other hand, 38.2\% (44.3\%) of galaxies in PM2GC
structures with $\sigma >400 km/s$ (500) are located in D4, 34.5\%
(35.0\%) in D3, 19.9\% (18.4\%) in D2 and 7.4\% (2.3\%) in D1,
indicating that most of PM2GC cluster galaxies are hosted in
rather high density regions.

At higher redshift, we have at our disposal both the field and cluster
galaxies from the ICBS, besides EDIsCS clusters. As in the local
Universe, the
local density ranges spanned are different in clusters and field:
$\Delta (\log <\Sigma>) =2.5dex$ versus $\Delta (\log
<\Sigma>)=4dex$.

As a consequence, the different local density distributions covered by
the different samples could be responsible for the different
trends we have detected in clusters and (general) field.

As we have seen in \S3.1 for the most massive groups in the PM2GC, 
in the  field at low-z 
the local density trends of the mass functions do not seem to be due to
the global environment.
To assess in detail the role of the global environment, in separate papers
we have analysed how the galaxy
stellar mass function depends on it.
Using the PM2GC and WINGS data, in Calvi et al. (2011b in preparation),
we find that the shape of the
mass distribution does not depend on whether galaxies belong to 
a galaxy system (group or cluster) or not. 
Indeed, we are not able to detect any
substantial difference in the shape of the cluster, group and  field
mass functions.
In \cite{global} we have carried out a similar
 analysis on galaxies located
at higher redshift ($z\sim 0.3-0.8$) exploiting the 
capabilities of the ICBS
 and those of EDisCS. 
Again our findings suggest an universality of the mass distribution
in different global environments (clusters, groups and field), 
at least for galaxies above the mass limit of our samples ($\log M_{\ast}/M_{\odot} \geq 10.2-10.5$).
Summarizing, in those works we do not detect a dependence of 
the mass distribution on the global environment. 
Hence, in our samples,  above the same mass we use here (that corresponds to the two mass limits), 
we detect differences among  mass distributions of galaxies located at 
different local densities but not in different global environments.

As a consequence, the evidence that the mass function does depend on
local density, raises an interesting question. Why does local density
play a more active role than global environment in shaping the mass
function, at both redshifts?

Recently, other evidence has been accumulated that the local environment is
more important than the global environment in shaping also
several of the main galaxy properties, not just the galaxy mass.
The results for two of these properties are most striking, and concern 
the red galaxy population, and the morphological types of galaxies.

In WINGS clusters, none of the characteristics of the colour-magnitude
red sequence (slope, scatter, luminous-to-faint ratio, blue
fraction, morphological mix on the red sequence) depend on global
cluster properties connected with cluster mass, such as cluster
velocity dispersion and X-ray luminosity. In contrast, {\it   all} of these
characteristics vary systematically with the local galaxy density \citep{valentinuzzi11}.

Also in WINGS, we have shown that the fractions of spiral, S0 and elliptical
galaxies do not vary systematically with cluster velocity dispersion 
and X-ray luminosity (Poggianti et al 2009), while a strong morphology-density
relation is present in WINGS as in any other sample (Fasano et al. in prep.).

In addition, \cite{balogh04}, analysing the colour distribution of bright ($M_r\leq 18$) galaxies 
in the local Universe ($z < 0.08$). 
found that 
 the red fraction of galaxies 
is a strong function of local
density, increasing from $\sim 10\%- 30\%$ of the population in the 
lowest density environments to $\sim 70\%$ at the highest
densities, while 
%The fraction of red galaxies 
within the virialized regions of clusters it
shows no significant
dependence on cluster velocity dispersion.

Also \cite{mart08} 
 found that bright galaxy properties do not clearly depend on cluster
mass for clusters more massive than
$M\sim 10^{14}M_{\odot}$, while they correlate with clustercentric distance.

Our results on global and local environments now 
allow and require a comparison with theoretical expectations, 
to understand whether
simulations predict a mass segregation with environment, both considering 
the initial and evolved halo
mass and the local density and how they predict the evolution with redshift 
as a function of the
environment. 

\section{CONCLUSIONS}
In this paper, we have tried to quantify the importance
of the local density in shaping the stellar galaxy mass function of galaxies 
located in different environments both at low and intermediate redshifts
taking directly into account also the cluster environment.

Our main conclusion is that at all redshifts and in all environment
local density plays a significant role in driving the mass
distribution.  

In the general field at low-z, local density influences the stellar
mass distribution both at low and high masses.  In the field at
high-z, the dependence exists at high masses, while our mass limit
does not allow us to inspect low masses.  On the other hand, in
clusters, the biggest differences are always confined at low masses.
If we perform a higher mass cut ($ \log M_{\ast}/M_{\odot}>10.1$ for
WINGS and $ \log M_{\ast}/M_{\odot}>10.4$ in EDisCS), every difference
in slope disappears.

We have found that not only the shape of the mass function depends on local 
density, but also the highest mass reached in each density bin:  very massive
galaxies (having excluded the cluster BCGs) seem to be located only in the highest density 
bin while they are
absent at lower densities
 (the so-called mass segregation).  

Comparing our results with those present in Calvi et al. (2011b in preparation)
and \cite{global} for the global environment, 
we conclude that local environment plays a much more visible role than 
global environment in shaping the stellar galaxy mass distribution.

\section*{Acknowledgments}
We thank the referee for useful comments. 
We would like to thank  the whole WINGS team for
stimulating discussions. We also would like to thanks Alfonso Arag\'on-Salamanca 
and Gabriella De Lucia whose suggestions
helped us to improve the paper.
BV and BMP acknowledge financial support from ASI contract I/016/07/0 
and ASI-INAF I/009/10/0. BV also acknowledges financial support from the
Fondazione Ing. Aldo Gini.

\label{lastpage}


\begin{thebibliography}{}
\bibitem[Baldry et al.(2006)]{baldry06} Baldry, I.~K., Balogh, 
M.~L., Bower, R.~G., Glazebrook, K., Nichol, R.~C., Bamford, S.~P., 
\& Budavari, T.\ 2006, \mnras, 373, 469 
\bibitem[Balogh et al.(2002)]{balogh02} Balogh, M.~L., et al.\ 
2002, \apj, 566, 123 
\bibitem[\protect\citeauthoryear{Balogh et al.}{2004}]{balogh04} 
Balogh M.~L., Baldry I.~K., Nichol R., Miller C., Bower R., Glazebrook K., 
2004, ApJ, 615, L101 
\bibitem[\protect\citeauthoryear{Bamford et 
al.}{2008}]{bamford08} Bamford S.~P., Rojas A.~L., Nichol R.~C., 
Miller C.~J., Wasserman L., Genovese C.~R., Freeman P.~E., 2008, MNRAS, 
391, 607 
\bibitem[{Bell} \& {de Jong}(2001)]{bj01} Bell, E.~F., \& de Jong, R.~S.\ 2001, \apj, 550, 212
\bibitem[Berta et al.(2006)]{berta06} Berta, S., et al.\ 2006, \aap, 451, 881 
\bibitem[Blanton et al.(2005)]{blanton05} Blanton, M.~R., Eisenstein, D., Hogg, D.~W., Schlegel, D.~J., \& Brinkmann, J.\ 2005, \apj, 629, 143 
\bibitem[\protect\citeauthoryear{Bolzonella, Miralles, \& Pell{\'o}}{2000}]{bolzonella00} Bolzonella M., Miralles J.-M., Pell{\'o} R., 2000, A\&A, 363, 476 
\bibitem[Bolzonella et al.(2010)]{bolzonella10} Bolzonella, M., et al.\ 2010, \aap, 524, A76 
\bibitem[{Brunner} \& {Lubin}(2000)]{brunner00} Brunner, R.~J., \& Lubin, L.~M.  2000, \aj, 120, 2851 
\bibitem[{Bruzual} \& {Charlot}(2003)]{bc03} Bruzual, G., \& Charlot, S.\ 2003, \mnras, 344, 1000 
\bibitem[Bundy et al.(2006)]{bundy06} Bundy, K., et al.\ 2006, 
\apj, 651, 120 
\bibitem[\protect\citeauthoryear{Calvi, Poggianti, 
\& Vulcani}{2011}]{rosa} Calvi R., Poggianti B.~M., Vulcani B., 2011, MNRAS, 416, 727 
\bibitem[\protect\citeauthoryear{Calvi et al.}{2011}]{rosa2} 
Calvi R., Poggianti B.~M., Fasano G., Vulcani B., 2011, arXiv, 
arXiv:1110.0802 



\bibitem[{Cava} {et al.}(2009)]{cava09} Cava, A., et al.\ 2009, \aap, 495, 707 
\bibitem[Chuter et al.(2011)]{chuter11} Chuter, R.~W., et al.\ 
2011, \mnras, 222 
\bibitem[Colbert et al.(2001)]{colbert01} Colbert, J.~W., 
Mulchaey, J.~S., \& Zabludoff, A.~I.\ 2001, \aj, 121, 808 
\bibitem[\protect\citeauthoryear{Colless et 
al.}{2001}]{colless01} Colless M., et al., 2001, MNRAS, 328, 1039 
\bibitem[\protect\citeauthoryear{De Lucia et al.}{2004}]{delucia04} De Lucia, G., et al.\ 
2004, \apjl, 610, L77 
\bibitem[\protect\citeauthoryear{De Lucia et 
al.}{2007}]{delucia07} De Lucia G., et al., 2007, MNRAS, 374, 809 
\bibitem[\protect\citeauthoryear{Dressler}{1980}]{dressler80} 
Dressler A., 1980, ApJ, 236, 351 
\bibitem[{Ebeling} {et~al.}(1996)]{ebeling96} beling, H., Voges, W., 
Bohringer, H., Edge, A.~C., Huchra, J.~P., 
\& Briel, U.~G. 1996, \mnras, 281, 799 
\bibitem[{Ebeling} {et~al.}(1998)]{ebeling98} Ebeling, H., Edge, 
A.~C., Bohringer, H., Allen, S.~W., Crawford, C.~S., Fabian, A.~C., Voges, 
W., \& Huchra, J.~P. 1998, \mnras, 301, 881 
\bibitem[{Ebeling} {et~al.}(2000)]{ebeling00} Ebeling, H., Edge, 
A.~C., Allen, S.~W., Crawford, C.~S., Fabian, A.~C., 
\& Huchra, J.~P. 2000, \mnras, 318, 333 
\bibitem[{Ellison} {et al.}(2009)]{ellison09} Ellison, S.~L., Simard, 
L., Cowan, N.~B., Baldry, I.~K., Patton, D.~R., 
\& McConnachie, A.~W.\ 2009, \mnras, 396, 1257 
\bibitem[{Fasano} {et al.}(2006)]{fasano06} Fasano, G., et al.\ 2006, \aap, 445, 805 
\bibitem[Fukugita et al.(1996)]{fukugita96} Fukugita, M., Ichikawa, T., Gunn, J.~E., Doi, M., Shimasaku, K., 
\& Schneider, D.~P.\ 1996, \aj, 111, 1748 
\bibitem[{Gehrels}(1986)]{gehrels86} Gehrels, N. 1986, \apj, 303, 336 
\bibitem[\protect\citeauthoryear{Gladders 
\& Yee}{2000}]{gladders00} Gladders M.~D., Yee H.~K.~C., 2000, AJ, 120, 2148 
\bibitem[Gr{\"u}tzbauch et al.(2011a)]{ruth11a} Gr{\"u}tzbauch, 
R., Conselice, C.~J., Varela, J., Bundy, K., Cooper, M.~C., Skibba, R., 
\& Willmer, C.~N.~A.\ 2011, \mnras, 411, 929 
\bibitem[Gr{\"u}tzbauch et al.(2011b)]{ruth11b} Gr{\"u}tzbauch, 
R., Chuter, R.~W., Conselice, C.~J., Bauer, A.~E., Bluck, A.~F.~L., 
Buitrago, F., \& Mortlock, A.\ 2011, \mnras, 412, 2361 
\bibitem[{Gonzalez} {et~al.}(2001)]{gonzales01} Gonzalez, A.~H., 
Zaritsky, D., Dalcanton, J.~J., \& Nelson, A.   2001, \apjs, 137, 117 
\bibitem[\protect\citeauthoryear{Gunawardhana et 
al.}{2011}]{Guna11} Gunawardhana M.~L.~P., et al., 2011, 
MNRAS, 415, 1647 
\bibitem[Haas et al.(2011)]{haas11} Haas, M.~R., Schaye, J., 
\& Jeeson-Daniel, A.\ 2011, arXiv:1103.0547 
\bibitem[{Halliday} {et~al.}(2004)]{halliday04} Halliday, C., et al.  2004, \aap, 427, 397 
\bibitem[Kauffmann et al.(2004)]{kauffmann04} Kauffmann, G., White, 
S.~D.~M., Heckman, T.~M., M{\'e}nard, B., Brinchmann, J., Charlot, S., 
Tremonti, C., \& Brinkmann, J.\ 2004, \mnras, 353, 713 
\bibitem[Kodama \& Bower(2001)]{kodama01} Kodama, T., \& Bower, R.~G.\ 2001, Astrophysics and Space Science Supplement, 277, 597 
\bibitem[Kroupa(2001)]{kr01} Kroupa, P.\ 2001, \mnras, 322, 231 
\bibitem[Liske et al.(2003)]{liske03} Liske, J., Lemon, D.~J., 
Driver, S.~P., Cross, N.~J.~G., \& Couch, W.~J.\ 2003, \mnras, 344, 307 
\bibitem[Maraston(2005)]{maraston05} Maraston, C.\ 2005, \mnras, 
362, 799 
\bibitem[\protect\citeauthoryear{Mart{\'{\i}}nez, Coenda, 
\& Muriel}{2008}]{mart08} Mart{\'{\i}}nez H.~J., Coenda V., Muriel H., 2008, MNRAS, 391, 585 
\bibitem[{Milvang-Jensen} {et~al.}(2008)]{milvang08} Milvang-Jensen, B., et al.  2008, \aap, 482, 419 
\bibitem[Moresco et 
al.(2010)]{moresco10} Moresco, M., et al.\ 2010, \aap, 524, A67 
\bibitem[Mouhcine et al.(2007)]{mouhcine07} Mouhcine, M., Baldry, 
I.~K., \& Bamford, S.~P.\ 2007, \mnras, 382, 801 
\bibitem[Muldrew et al.(2011)]{muldrew11} Muldrew, S.~I., Croton, 
D.~J., Skibba, R.~A., et al.\ 2011, arXiv:1109.6328 
\bibitem[Pasquali et al.(2009)]{pasquali09} Pasquali, A., van den 
Bosch, F.~C., Mo, H.~J., Yang, X., \& Somerville, R.\ 2009, \mnras, 394, 38 
\bibitem[Pell{\'o} {et al.}(2009)]{pello09} Pell{\'o}, R., et al.\ 2009, \aap, 508, 1173 
\bibitem[Peng {et al.}(2010)]{peng10} Peng, Y.-j., et al.\ 2010, \apj, 721, 193 
\bibitem[{Poggianti}(1997)]{poggianti97} Poggianti, B.~M.\ 1997, \aaps, 122, 399 
\bibitem[{Poggianti} {et~al.}(2008)]{poggianti08} Poggianti, B.~M., et al.  2008, \apj, 684, 888 
\bibitem[Roberts et al.(2007)]{roberts07} Roberts, S., Davies, 
J., Sabatini, S., Auld, R., \& Smith, R.\ 2007, \mnras, 379, 1053 
\bibitem[{Rudnick} {et al.}(2001)]{rudnick01} Rudnick, G., et al. 2001, \aj, 122, 2205 
\bibitem[{Rudnick} {et al.}(2003)]{rudnick03} Rudnick, G., et al. 2003, \apj, 599, 847 
\bibitem[{Rudnick} {et~al.}(2009)]{rudnick09} Rudnick, G., et al. 2009, \apj, 700, 1559 
\bibitem[Scodeggio et al.(2009)]{scodeggio09} Scodeggio, M., et al.\ 2009, \aap, 501, 21 
\bibitem[Scoville et al.(2007)]{scoville07} Scoville, N., et al.\ 
2007, \apjs, 172, 1 
\bibitem[{Taylor} {et al.}(2009)]{taylor09} Taylor, E.~N., et al.\ 
2009, \apjs, 183, 295 
\bibitem[{Valentinuzzi} {et~al.}(2009)]{valentinuzzi09} Valentinuzzi, T., et al. 2009, \aap, 501, 851 
\bibitem[\protect\citeauthoryear{van der Wel}{2008}]{vanderwel08} 
van der Wel A., 2008, ApJ, 675, L13 
\bibitem[\protect\citeauthoryear{Valentinuzzi et 
al.}{2011}]{valentinuzzi11} Valentinuzzi T., et al., 2011, arXiv, 
arXiv:1109.4011 
\bibitem[{Varela} {et al.}(2009)]{varela09} Varela, J., et al.\ 2009, \aap, 497, 667 
\bibitem[\protect\citeauthoryear{von der Linden et 
al.}{2010}]{vonderlinden10} von der Linden A., Wild V., Kauffmann G., 
White S.~D.~M., Weinmann S., 2010, MNRAS, 404, 1231 
\bibitem[Vulcani et al.(2010)]{vulcani10} Vulcani, B., Poggianti, B.~M., Finn, R.~A., Rudnick, G., Desai, V., 
\& Bamford, S.\ 2010, \apjl, 710, L1 
\bibitem[\protect\citeauthoryear{Vulcani et 
al.}{2011}]{morph} Vulcani B., et al., 2011, MNRAS, 412, 246 
\bibitem[\protect\citeauthoryear{Vulcani et 
al.}{2011b}]{global} Vulcani B., et al., 2011, A\&A, submitted
\bibitem[Wolf et al.(2009)]{wolf09} Wolf, C., et al.\ 2009, 
\mnras, 393, 1302 
\bibitem[Weinmann et al.(2009)]{weinmann09} Weinmann, S.~M., 
Kauffmann, G., van den Bosch, F.~C., Pasquali, A., McIntosh, D.~H., Mo, H., 
Yang, X., \& Guo, Y.\ 2009, \mnras, 394, 1213 
\bibitem[{White} {et al.}(2005)]{white05} White, S.~D.~M., et al.\ 2005, \aap, 444, 365 
\bibitem[\protect\citeauthoryear{York et al.}{2000}]{york00} 
York D.~G., et al., 2000, AJ, 120, 1579 
\bibitem[Zandivarez 
\& Martinez(2010)]{zandivarez10} Zandivarez, A., \& Martinez, H.~J.\ 2010, arXiv:1012.3445 
\end{thebibliography}
\end{document}